\documentclass[pre]{revtex4}%
\usepackage{amsfonts}
\usepackage{amsmath}
\usepackage{amssymb}
\usepackage{graphicx}%
\usepackage{epsfig}
\usepackage{epstopdf}
\providecommand{\U}[1]{\protect\rule{.1in}{.1in}}

\begin{document}
\title{The Thermodynamic Behaviors and Glass Transition on the Surface/Thin Film of
An Ising Spin Model on Recursive Lattice}
\author{Ran Huang \footnote{Correspondence to: ranhuang@sjtu.edu.cn}}
\affiliation{School of Chemistry and Chemical Engineering, Shanghai Jiao Tong University, Shanghai 200240, China}
\author{Purushottam D. Gujrati \footnote{Co-correspondence to: pdg@uakron.edu}}
\affiliation{The Department of Physics, The University of Akron, Akron, OH\ 44325}

\begin{abstract}
A quasi 2-dimensional recursive lattice formed by planar elements have been
designed to investigate the surface thermodynamics of Ising spin glass system
with the aim to study the metastability of supercooled liquids and the ideal
glass transition. The lattice is constructed as a hybrid of partial Husimi
lattice representing the bulk and 1D single bonds representing the surface. The recursive properties of the
lattices were adopted to achieve exact calculations. The model has an anti-ferromagnetic interaction to give
rise to an ordered phase identified as crystal, and a metastable solution representing the amorphous/metastable phase. Interactions between
particles farther away than the nearest neighbor distance are taken into
consideration. Free energy and entropy of the
ideal crystal and supercooled liquid state of the model on the surface are
calculated by the partial partition function. By analyzing the free energies
and entropies of the crystal and supercooled liquid state, we are able to
identify the melting transition and the second order ideal glass transition on
the surface. The results show
that due to the coordination number change, the transition temperature on the
surface decreases significantly compared to the bulk system. Our calculation
agrees with experimental and simulation results on the thermodynamics
of surfaces and thin films conducted by others.

\end{abstract}
\date{\today}
\maketitle

\section{INTRODUCTION}

Glass transition on surface/interface/thin film has drawn intensive interests in
the last two decades for two reasons. Firstly, the importance of surface
and thin film in materials science and engineering requires a better
understanding of its dynamic and thermodynamic properties. Secondly, the confined geometry is a good
approach to understand the mysterious glass transition itself, especially to explore the dynamic
properties within the thin film geometry. Numerous works of
both experimental and simulation/calculation approaches have been done [1-23]
on glass transition on surface/thin film.

Keddie and co-workers firstly investigated the supported thin
film of PS by ellipsometric measurements \cite{1}. They prepared several PS
films supported by silicon wafers with thickness from  $100%
\operatorname{\text{\AA}}%
$ to $3000%
\operatorname{\text{\AA}}%
$. The measurements indicated the deduction of the $T_{\text{g}}$ with thickness under 400$%
\operatorname{\text{\AA}}%
$. An empirical relationship between thickness and $T_{\text{g}}$ was given as:%

\begin{equation}
T_{\text{g}}=T_{\text{g}}^{\text{Bulk}}[1-(\frac{\alpha}{h})^{\delta}],
\label{TG_reduction}%
\end{equation}
where $T_{\text{g}}^{\text{Bulk}}$ is the glass transition temperature of the bulk
PS. The $\alpha$ and $\delta$ are the parameters fitted to be $32%
\operatorname{\text{\AA}}%
$ and $1.8$ respectively, $h$ is the thickness of the film. Following Keddie's
work, many researches have been done on different supported thin films of various polymers [2-3] by
different characterization methods, such as X-ray reflectivity, positron annihilation, and dielectric \cite{4,5,6}. Most results have demonstrated the same phenomenon that
for liner polymer the $T_{\text{g}}$ decreases with the thickness of films.
However the supported film has a considerable film-substrate interaction,
which makes the conclusion controversial. Strong attractive interaction
between the substrate and thin film may increase the $T_{\text{g}}$ of thin
film above the bulk $T_{\text{g}}$ [6]. van Zanten et al. measured
$T_{\text{g}}$ of poly-2-vinylpyridine on oxide-coated Si substrates, and
found it increase by  $50$K than the bulk $T_{\text{g}}$, for a $77%
\operatorname{\text{\AA}}%
$ film [7]. Forrest and co-workers have done pioneer works in measuring the
$T_{\text{g}}$ of free-standing thin films [8-10]. They measured the
$T_{\text{g}}$ of free-standing PS films with thickness from $200$ to $2000%
\operatorname{\text{\AA}}%
$ and different molecular weights by Brillouin Light Scattering and
transmission ellipsometry. Their results showed that the $T_{\text{g}}$
decreases with the thickness of PS thin film with a much larger magnitude: for
example, the $T_{\text{g}}$ of $200%
\operatorname{\text{\AA}}%
$ film with \textit{Mw} within $120$K to $378$K reduces by 70K below the
$T_{\text{g}}^{\text{Bulk}}$, while this magnitude is around $10$K for
supported films. The empirical equation (Eq.\ref{TG_reduction}) derived for
supported film still holds for the low \textit{Mw} free-standing films. With
$\delta =1.8$, the parameter $\alpha $ was found to be $78%
\operatorname{\text{\AA}}%
$ which is twice of it found for supported films. 

Other than the experimental work, computer simulation and
calculation have also been developed to investigate the glass transition on
surface/thin film, and most of them are for polymer systems [11-23]. Molecular Dynamics (MD) and Monte Carlo (MC) method were usually employed with various modelings, and confirmed the
$T_{\text{g}}$ decrease with the thickness reduction for both supported and free
standing film, or $T_{\text{g}}$ increase in some particular substrate-film cases. The Eq.\ref{TG_reduction} derived from experiments can also be validated by simulations, nevertheless the explanation for the mechanism of $T_{\text{g}}$ reduction is still a matter of
debate. Most MD simulations verified the experimental observation that for
supported film, a strong substrate-film interaction will increase the
$T_{\text{g}}$ above the value in the bulk, while the weak substrate-film
interaction will lead $T_{\text{g}}$ reduction, and free-standing film shows a
much larger $T_{\text{g}}$ reduction than supported film with weak
substrate-film interaction [12]. Mattice and co-workers firstly reported the MC
simulation in this field [13, 14]. de Pablo et.al. reported a MC
simulation on free-standing films of both linear and cyclic polymeric chains
[23]. Basically the $T_{\text{g}}$ reduction with smaller thickness was
confirmed in the above investigations. In de Pablo and co-workers' work, the
relation between $T_{\text{g}}$ and $T_{\text{g}}^{\text{Bulk}}$ can also be
fitted to be Eq. \ref{TG_reduction} with slight difference on the parameters fitting.

Although the glass transition is not a unique property of polymers, most
simulation works were modeled for polymer systems, while the works on small
molecule systems are very rare. Meanwhile, Ising spins glass has also been
widely utilized to study the glass transition [24-37]. By different lattices
adopted and interactions setup the Ising model is capable to describe various
systems, such as gas, liquid, crystal and glass, and consequently the phase
transitions, like melting and glass transition [24]. However, very few of the
efforts were on surface/thin film glass transition of Ising spin glass. This
field had been explored by Gujrati et al. by applying Ising model on a
modified Bethe lattice to describe the thermodynamics of polymer systems near
surface [38-42]. In this paper we follow the similar method to study the
glass transition of Ising spin glass on the surface/thin film on a specially
constructed recursive lattice.

\section{SURFACE RECURSIVE LATTICE (SRL) GEOMETRY}

\begin{figure}
[ptb]
\begin{center}
\includegraphics[width=0.8\textwidth]
{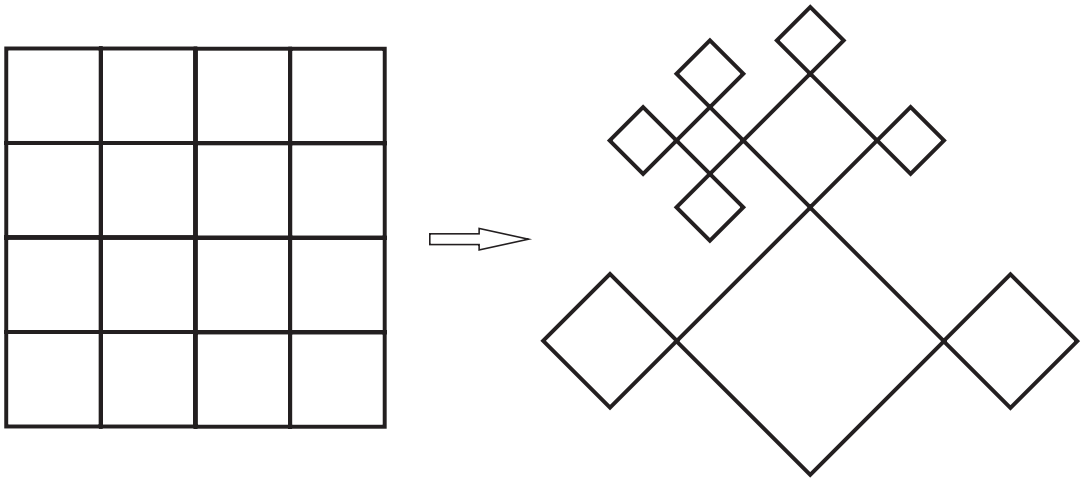}%
\caption{A regular square lattice and its analog of a recursive lattice
(Husimi lattice).}%
\label{fig1}%
\end{center}
\end{figure}

Except in some rare cases [43-45], a many-body system with interactions on a
regular lattice is difficult to be solved exactly because of the complexity
involved with treating the combinatorics generated by the interaction terms in
the Hamiltonian when summing over all states. The mean-field approximation is
commonly adopted to solve this problem, e.g. the Flory model of semiflexible
polymers [46, 47]. On the other hand, recursive lattices enable us to take the
explicit treatment of combinatorics on these lattices and no approximation is
necessary [25-28, 48]. The recursive lattice is chosen to have the same
coordination number as the regular lattice it is designed to describe. As
usual, the coordination number $q$ is the number of nearest-neighbor sites of
a site. A typical recursive
lattices, the Husimi lattice, as the analogs of the 2-D lattice is shown in
Fig. \ref{fig1}. For the bulk system, the recursive lattice calculations have
been demonstrated to be highly reliable approximations to regular lattices [48-50].%

In this work we construct a recursive lattice to describe the
surface/thin film. This surface recusive lattice (SRL) is to mimic the 2D
case, i.e. the 2D bulk with 1D surfaces. The SRL is integrated of square units
representing the bulk and single bond representing the surface, the structure
is made to have the same coordination numbers with regular 2D square lattice.
Ising spins will be applied on the lattices to represent a small molecule
system. The exact calculation will be drived and the solutions will be discussed.

\subsection{Construction of SRL}

\subsubsection{Structure}

\begin{figure}
[ptb]
\begin{center}
\includegraphics[width=0.8\textwidth]
{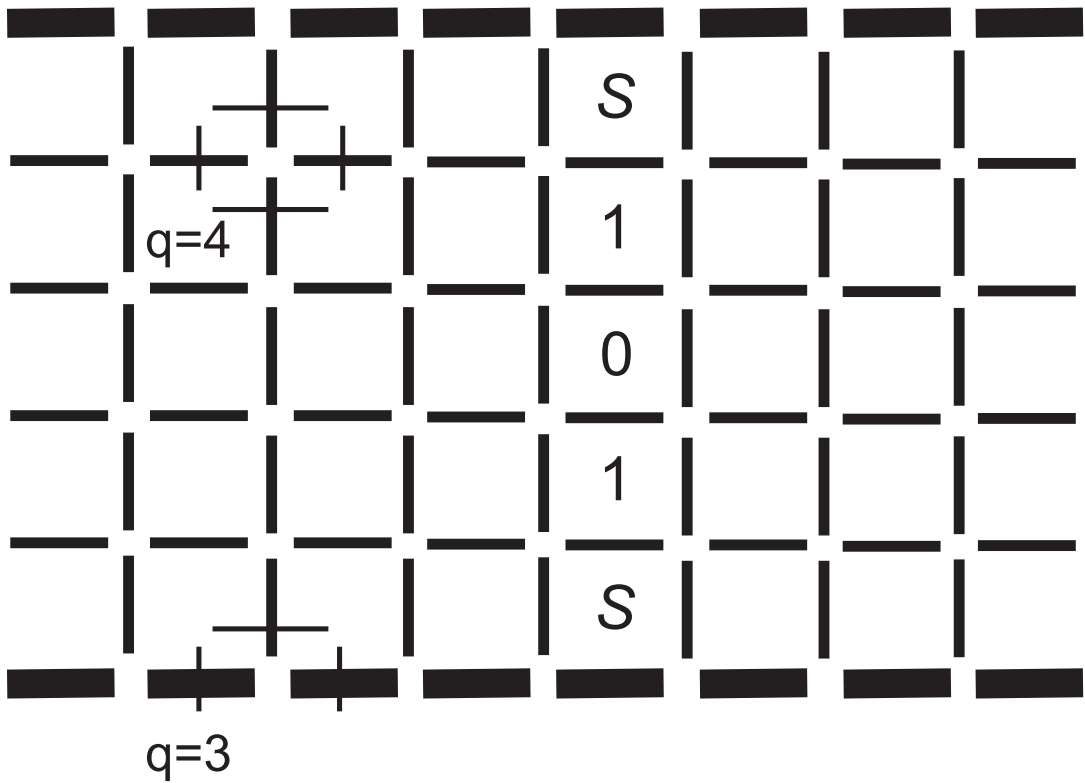}%
\caption{A regular 2D square lattice of a thin film with thickness = $5.$}%
\label{fig2}%
\end{center}
\end{figure}

To approximate the regular lattice due to the same coordination number, we
firstly need to figure out the coordination number and interactions of a
regular lattice of surface/thin film to construct the SRL to represent the
surface/thin film. The Fig. \ref{fig2} shows a regular 2D square lattice of a
thin film with thickness equals to 5 and the thick line is the surface. (In
all figures of this paper the thick bond represents surface bond.) Here we
label the central layer to be the $0$th level, and the layer next to $0$th
layer is the level $1$st. The surface layer is labeled as level $S$. The sites
on the surface have coordination number of $3$, while inside the bulk the
coordination number is $4$. That is, a hybrid recursive lattice with $q=3$ and
$4$ is required to describe the surface/thin film.%

\begin{figure}
[ptb]
\begin{center}
\includegraphics[width=0.8\textwidth]
{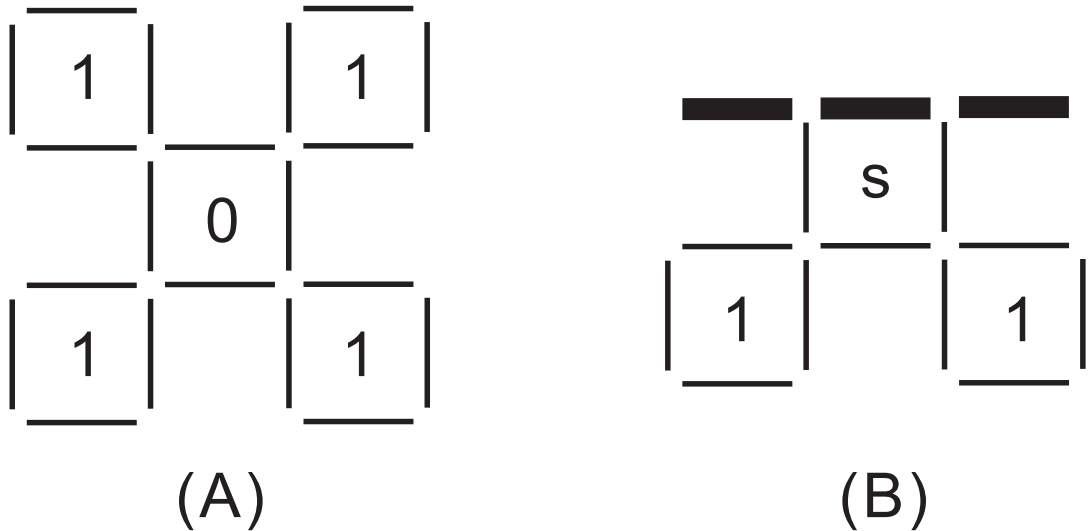}%
\caption{The basic unit in the bulk and on the surface of a regular 2D square
lattice}%
\label{fig3}%
\end{center}
\end{figure}

\begin{figure}
[ptb]
\begin{center}
\includegraphics[width=0.8\textwidth]
{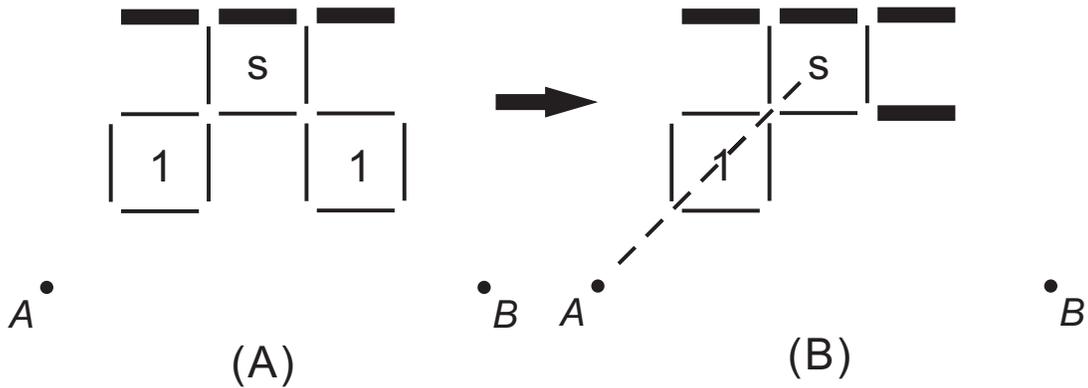}%
\caption{The modification of surface square unit. The bulk square on lower
right corner is replaced by a single bond. In this way the modified structure
is symmetrical to an imaginary axis links point A and the center of square
unit S.}%
\label{fig4}%
\end{center}
\end{figure}

Now we take out the basic units inside the bulk and on the surface (Fig.
\ref{fig3}) to construct a recursive lattice. Inside the bulk we simply have
Husimi lattice of $q=4$. For the surface structure, a single bond representing
in the surface unit links on a square unit then we can have sites with $q=3$.
However this unit cannot be simply adopted to construct a recursive structure,
because the recursive calculation technique (will be discussed later) requires
an origin point to which the entire tree is symmetrical. For the unit shown in
Fig. \ref{fig4}a, wherever we determine the origin is (point A or B), the unit
is not symmetrical to that point. We modify the surface unit by replacing one
square by an artificial single bond at one lower corner. Although this unit
differs from the regular lattice we want to approximate, the coordination
numbers on the surface square still accord to our design and the calculation
in the following sections will show that this approximation is practical. The
modification is shown in Fig. \ref{fig4}, the modified unit (Fig. \ref{fig4}b)
is then symmetrical to the point A.%

%

\begin{figure}
[ptb]
\begin{center}
\includegraphics[width=0.8\textwidth]
{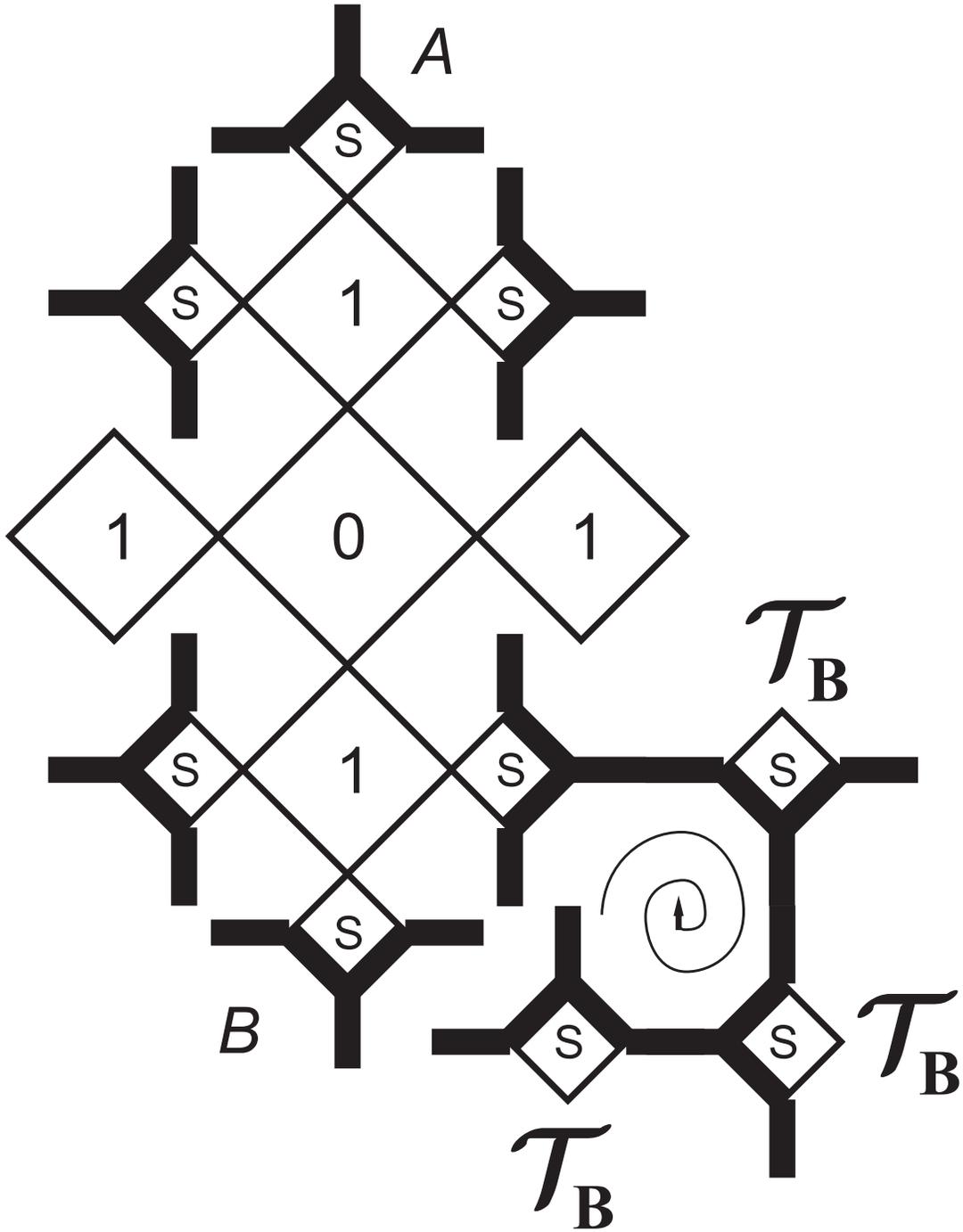}%
\caption{A bulk tree structure in surface recursive lattice with thickness =
5.}%
\label{fig5}%
\end{center}
\end{figure}

Then we can put the bulk and surface units together to construct a recursive
lattice to approximate the regular thin film lattice. The structure of a
thickness $=5$ lattice is shown in Fig. \ref{fig5}.%

\begin{figure}
[ptb]
\begin{center}
\includegraphics[width=0.8\textwidth]
{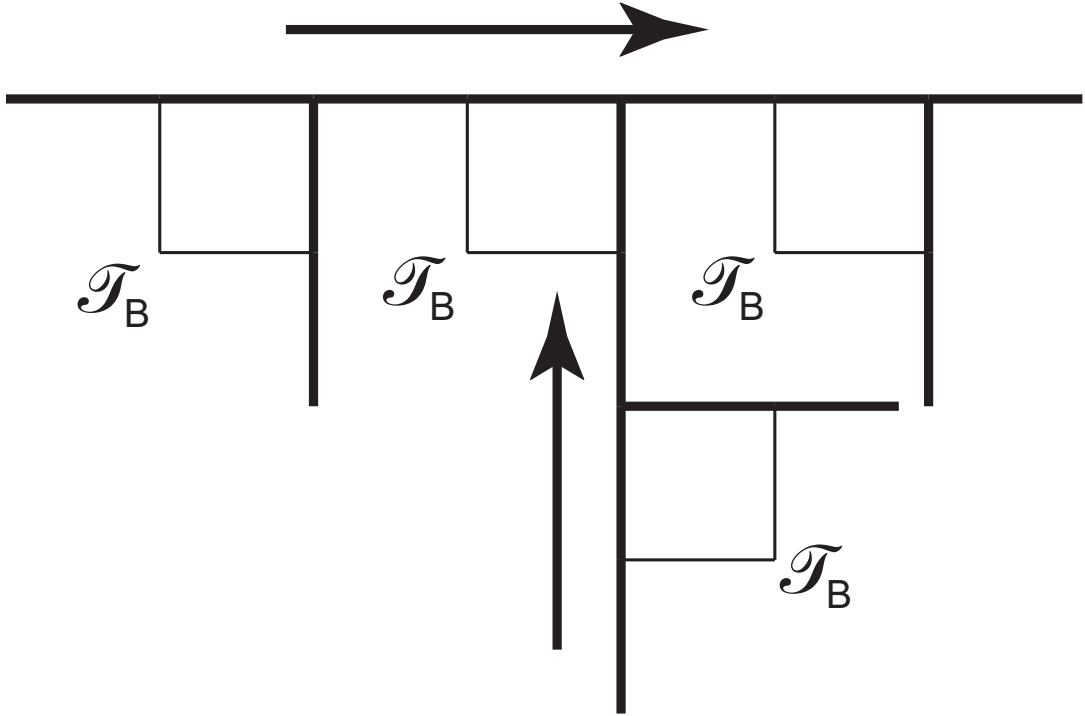}%
\caption{The stretched view and the contribution direction of one surface in
SRL.}%
\label{fig6}%
\end{center}
\end{figure}

From surface A to B, we have a finite bulk portion with a thickness $=5$, and
these five layers are labeled as $S$, $1$, and $0$ like the regular film
lattice in Fig. \ref{fig2}. The thick single bond representing surface links
two identical bulk trees. In this example, one bulk tree \textit{T}%
$_{\text{\textbf{B}}}$ is linked overall with another $36$ trees (only one
bulk tree is drawn in above Fig. \ref{fig5}), while they are all identical and
independent with each other except a single surface bond connection.
Recursively, each of these $36$ trees is also linked with another $35$ trees
by single surface bonds. This lattice is an infinite tree integrated by the
finite size bulk portions and infinitely large surfaces. If we look at the
surface integrated by thick surface bonds indicated by the arrow at the right
lower corner in Fig. \ref{fig5}, we can see the surface going through the bulk
trees \textit{T}$_{\text{\textbf{B}}}$ is infinitely large with a coordination
number 3 everywhere. A stretched surface is shown in Fig. \ref{fig6} to
provide a more obvious view of the surface. Comparing to the regular lattice
surface, the main difference is that one surface also receives thermodynamic
contributions from other surface structures, which is caused by the surface
unit modification. But this is an approximation we have to take to achieve the
recursive calculation. Although we have infinite number of bulk trees and
surfaces in this lattice, since they are independent and identical, it does
not impact the thermodynamic properties of a local region we are going to investigate.%

\subsubsection{The sites labeling on bulk and surface units}

We label each site in a square as shown in Fig. \ref{fig7}:%

\begin{figure}
[ptb]
\begin{center}
\includegraphics[width=0.8\textwidth]
{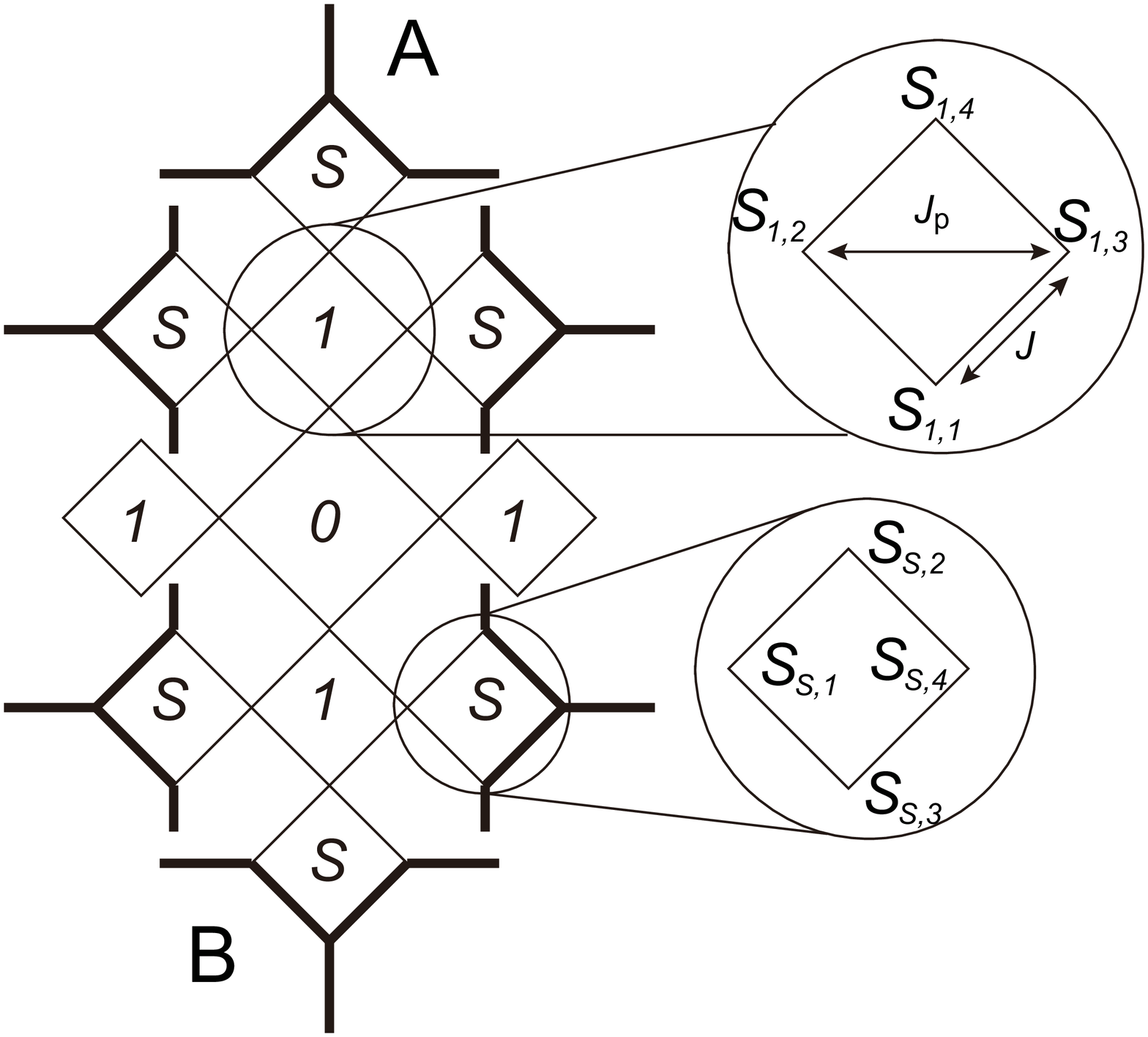}%
\caption{Sites index on a square unit in SRL.}%
\label{fig7}%
\end{center}
\end{figure}

The site in a square unit is labeled as \textit{S}$_{a,b}$, where $a$ is the
index of the square unit in the bulk tree, $b$ is the index of site.
Therefore, one site \textit{S}$_{a,b}$ actually has two labels depending on
which square unit we refer to. For example the site \textit{S}$_{a,4}$ is also
\textit{S}$_{(a+1),1}$ in the higher level unit. When we are only focusing on
one unit, the site \textit{S}$_{a,b}$ can be denoted as \textit{S}$_{b}$ for
convenience. In a square unit, the base site is determined to be the site
closest to the bulk tree origin (The $0$th square) and labeled as
\textit{S}$_{a,1}$. Therefore, in the upper enlarged circle in Fig.
\ref{fig7}, the base site is the lowest site, while in the lower enlarged
circle, the base site is the left one. The labeling in the origin unit is a
special case which will be discussed later. The interactions are also sampled
in the enlarged circle: $J$ is the interaction between the nearest sites,
$J_{\text{p}}$ is the interaction between the second-nearest sites, i.e. the
diagonal sites.

\begin{figure}
[ptb]
\begin{center}
\includegraphics[width=0.8\textwidth]
{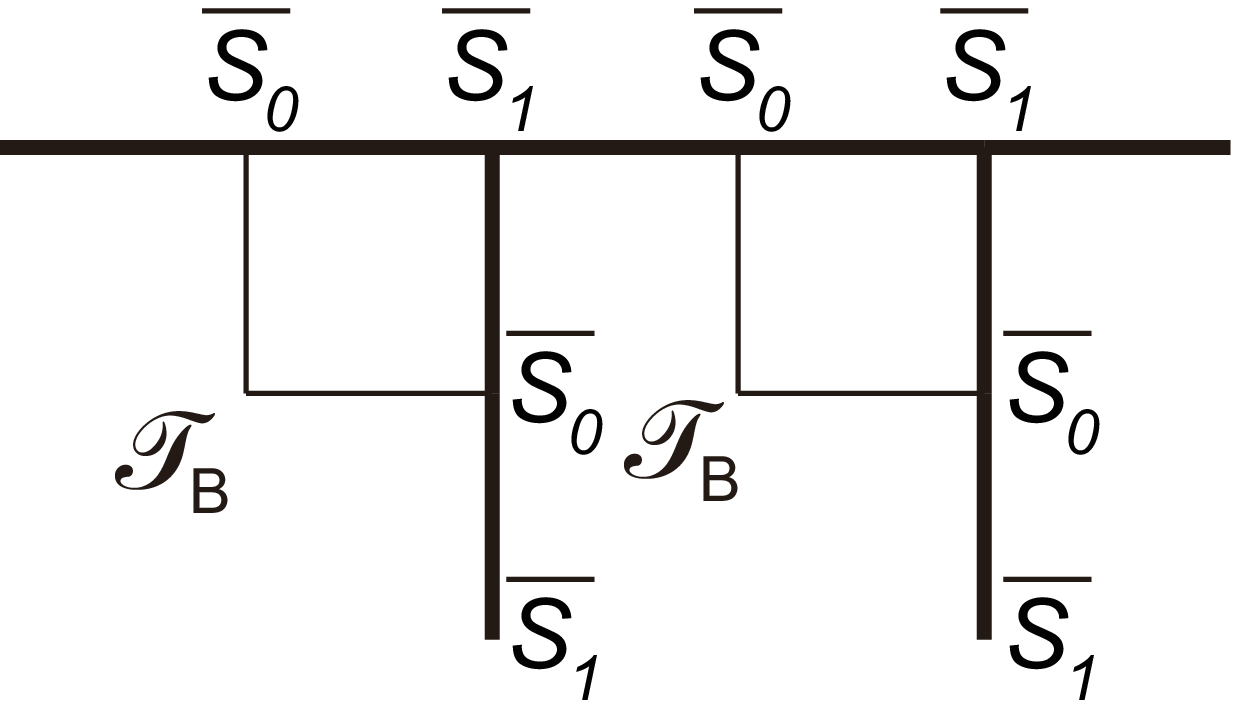}%
\caption{The surface sites labeling in SRL.}%
\label{fig8}%
\end{center}
\end{figure}

For surface calculation, we label the surface sites as shown in Fig.
\ref{fig8}. Due to our calculation method, only two labels are necessary in
surface labeling. The site close to the bulk site is labeled as $\overline
{S_{0}}$, and the site being diagonal to the bulk site is labeled as
$\overline{S_{1}}$.%

1.3) The interactions and energy in bulk square unit

We consider four kinds of interactions in our model: $J$ the interaction
energy between the nearest sites, $J_{\text{p}}$ the interaction energy
between the second nearest sites, $J\prime$ the interaction energy of three
sites (triplet), and $J"$ the interaction energy of four sites (quadruplet).

We introduce the following variables to count the interactions and magnetic
fields of one square unit:%

\begin{align*}
A_{\text{nb}}  &  =S_{1}\times S_{2}+S_{1}\times S_{3}+S_{2}\times S_{4}%
+S_{3}\times S_{4};\\
A_{\text{d}}  &  =S_{1}\times S_{4}+S_{2}\times S_{3};\\
A_{\text{tri}}  &  =S_{1}\times S_{2}\times S_{3}+S_{1}\times S_{2}\times
S_{4}+S_{1}\times S_{3}\times S_{4}+S_{2}\times S_{3}\times S_{4};\\
A_{\text{qua}}  &  =S_{1}\times S_{2}\times S_{3}\times S_{4};\\
A_{\text{mag}}  &  =S_{2}+S_{3}+S_{4}.
\end{align*}

Where $S_{m}$ is an Ising spin and has the value of $+1$ or $-1$. Note that
the base site $S_{1}$ is not included in the magnetic term $A_{\text{mag}}$,
because the calculation has a recursive fashion that the energy of one unit is
included in the contribution to the next level's unit, and the base site of
one unit will be counted as the top site in the next level's energy equation,
thus we need to exclude it here to avoid the double counting.

Then, for a particular cell $\alpha $ with a certain spin conformation
$\Gamma $, the energy of the Ising model in one cell is:%

\begin{equation}
e_{\alpha}=-J\cdot A_{\text{nb}}(\Gamma )-J_{\text{p}}\cdot A_{\text{d}%
}(\Gamma )-J\prime\cdot A_{\text{tri}}(\Gamma )-J"\cdot A_{\text{qua}%
}(\Gamma )-H\cdot A_{\text{mag}}(\Gamma ). \label{e_alpha}%
\end{equation}

The total energy of the Ising model on SRL is the sum of the energy of all cells:%

\begin{equation}
E_{total}=\underset{\alpha}{%
{\displaystyle\sum}
}e_{\alpha}(\Gamma ). \label{Model_Energy}%
\end{equation}

In this work we set $J$ to be negative to adopt the the anti-ferromagnetic
Ising model, the neighbor spins preferring different states corresponds to the
lowest energy stable state, i.e. the ideal ordered crystal conformation, which
have the largest weight comparing to other conformations at the same
temperature, while the neighbor spins of the same states are the most unstable
conformations with the highest energy, which represent the disordered phase.

Then the Boltzmann weight $\omega(\Gamma )$ of state $\Gamma $ is given by%

\begin{equation}
\omega(\Gamma )=\exp[-\beta(e(\Gamma)-HS_{0})], \label{weights}%
\end{equation}

where $\beta $ $=1/k_{\text{B}}T$ is the inverse temperature, and here we
set the Boltzmann constant $k_{\text{B}}=1$ to make the temperature to be
generalized in the unit of energy.

Then the partition function of a finite lattice with $ \alpha $ cells is the sum
over products over the Boltzmann weight:%

\begin{equation}
Z(T)=\underset{\{S=\pm1\}}{%
{\displaystyle\sum}
}[\underset{\alpha}{%
{\displaystyle\prod}
}\omega_{\alpha}(\Gamma )]. \label{PF}%
\end{equation}

\subsubsection{The interactions and energy on surface}

In Fig. \ref{fig6}, the single bond unit and the square unit alternatively
appear on the surface structure. The interactions and energy of the surface
square unit is similar to the bulk square unit as we discussed in the previous
section. However, here we would like to assign a different nearest neighbor
interaction $\overline{J}$ on the surface bond, to distinguish it with $J$ in
the bulk. This enables us to investigate more complex surface properties. And
similarly we may also have a different diagonal interaction ($\overline{J_{p}%
}$), triplet interaction ($\overline{J\prime}$), quadruplet interaction
($\overline{J"}$) and magnetic field ($\overline{H}$):%

\begin{equation}
e_{\alpha}=-\overline{J}\cdot(S_{2}\times S_{4}+S_{3}\times S_{4}%
)-J\cdot(S_{1}\times S_{2}+S_{1}\times S_{3})-\overline{J_{\text{p}}}\cdot
A_{\text{d}}(\Gamma )-\overline{J\prime}\cdot A_{\text{tri}}(\Gamma
)-\overline{J"}\cdot A_{\text{qua}}(\Gamma )-\overline{H}\cdot A_{\text{mag}%
}(\Gamma ). \label{surface_e_alpha}%
\end{equation}

On the surface single bond unit, the interaction is much simpler since there
is only one nearest neighbor interaction:%

\begin{equation}
e=-\overline{J}\cdot(\overline{S_{0}}\times\overline{S_{1}})-\overline{H}%
\cdot\overline{S_{1}}. \label{single_bond_e}%
\end{equation}

Similar to the role of base site in bulk calculation, here the magnetic field
of $\overline{S_{0}}$ is not included in the second term because that site
will be counted in the next level's energy equation. The `level', in this
statement, is indexed by taking the direction indicated by the arrow in Fig.
\ref{fig6}, and employing an imaginary origin point infinitely far away from
the region we concern. Since the surface is infinitely large, the selection of
`origin point' does not affect us to achieve the solutions on surface.

By the setup of interaction energy parameters, we can simulate various systems
with particular interactions and energy to study their thermodynamic
properties and phase transition with the determination of the partition
function. The effect of energy parameters will be discussed in section IV. We
will generally set $J=-1$ to determine the temperature scale for our
antiferromagnetic model. The solution based on $J=-1$ and all other parameters
to be $0$ is called the \emph{reference model}.

\subsection{General recursive calculation technique}

\begin{figure}
[ptb]
\begin{center}
\includegraphics[width=0.8\textwidth]
{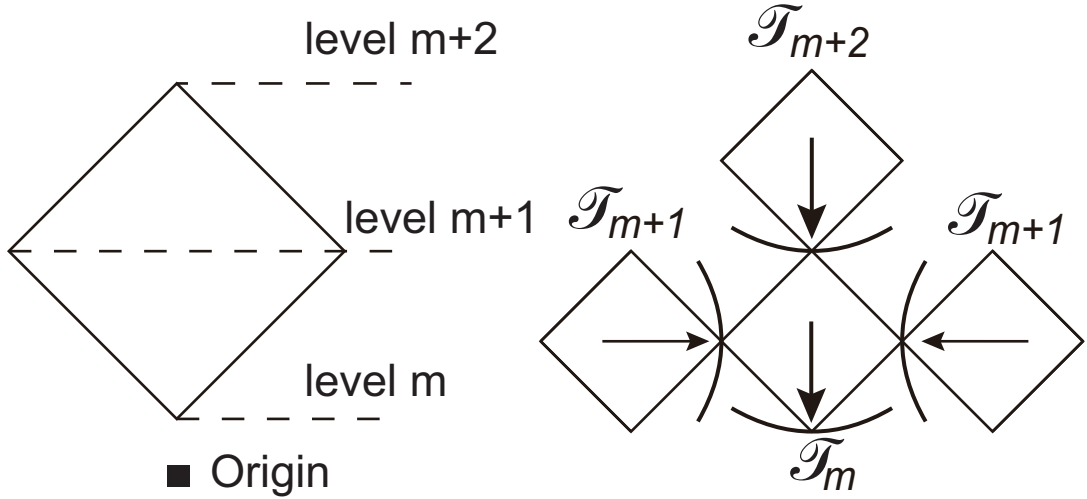}%
\caption{The levels on a square unit and the sub-tree contribution.}%
\label{fig9}%
\end{center}
\end{figure}

To discuss the calculation of our model, we first need to introduce the
concept of sub-tree contributions. In the finite bulk tree we have an origin
at the center of the tree. For each square unit, the base site is the closest
site to the origin, and there are three sub-trees coming from the other three
sites. The sub-trees could either be three identical portions of the bulk
tree, or three surface trees linked by the single surface bond if it is the
square unit on the surface. We label the levels in one unit and show the
sub-tree contributions in Fig. \ref{fig9}.%

By this index, the sites on the level $m$ is represented as $S_{m}$. The
sub-tree going to the site on level $m$ is marked as \textit{T}$_{m}$. In this
way, we can introduce the partial partition function (PPF) $Z_{m+1}(S_{m+1})$
at the level $m+1$ to represent the contribution of the branch \textit{T}%
$_{m+1}$ with a certain state of spin $S_{m+1}$ as its base site to the lower
level $m$, and \textit{T}$_{m}$\ to the lower level $m-1$, etc. Therefore the
PPF $Z_{m}(S_{m})$ at level $m$ is a function of the PPFs $Z_{m+1}(S_{m+1})$
and $Z_{m+2}(S_{m+2})$\ at higher level $m+1$ and\ $m+2$ and the local weight.
Then we can start from the highest level, count the contributions of sub-trees
on each level and recursively go to the lower level unit the origin point,
where it gains the contribution of the entire lattice, and the thermodynamics
of the entire system can be obtained by the partition function. The PPF of a
sub-tree \textit{T}$_{m}$\ $Z_{m}(\pm)$\ is the sum of the configurations with
the spin $S_{m}=\pm1$ over the products of the PPFs on higher levels with the
local weight $\omega(\Gamma)$. For a square unit, $Z_{m}(+)$ has $8$ terms
with $S_{m}=+1$ and $Z_{m}(-)$ is the sum of the other $8$ configurations with
$S_{m}=-1$:%

\begin{align}
Z_{m}(+)  &  =\underset{\Gamma=1}{\overset{8}{\sum}}[\underset{\{m+1\}}%
{\overset{2}{%
{\displaystyle\prod}
}}Z_{m+1}(S_{m+1})]Z_{m+2}(S_{m+2})w(\Gamma_{S_{m}=1}), \label{PPF+_Recursion}%
\\
Z_{m}(-)  &  =\underset{\Gamma=9}{\overset{8}{\sum}}[\underset{\{m+1\}}%
{\overset{2}{%
{\displaystyle\prod}
}}Z_{m+1}(S_{m+1})]Z_{m+2}(S_{m+2})w(\Gamma_{S_{m}=-1}),
\label{PPF-_Recursion}%
\end{align}

Depending on the spin state ($S_{0}=\pm1$) of the origin site, the two
sub-trees' contributions to the whole system are:
\[
Z_{0}^{2}(+)\exp(\beta H)\text{ or }Z_{0}^{2}(-)\exp(-\beta H).
\]

where the magnetic term is to count the contribution of origin site which is
not contained in Eq.\ref{e_alpha}. The partition function $Z_{0}$ of the
entire lattice can be accounted by the contributions to the origin site:%

\begin{equation}
Z_{0}=Z_{0}^{2}(+)\exp(\beta H)\text{+}Z_{0}^{2}(-)\exp(-\beta H).
\label{PF_of_PPFs}%
\end{equation}

The first term is to count the conformation with the origin site spin
$S_{0}=+1$, the square of PPF represents two sub-trees and the weight of the
origin spin itself is counted as $\exp(\beta H)$, similiarly for the
$S_{0}=-1$ situation in the second term.

Or, if the origin is defined to be a square unit, the partition function is%

\[
Z_{0}=\overset{16}{\underset{\Gamma}{%
{\displaystyle\sum}
}}(\overset{4}{\underset{i}{%
{\displaystyle\prod}
}}Z_{i}(S_{i})\cdot\omega(\Gamma))
\]

where $S_{i}$ is one of the four spins on the origin square, $Z_{i}(S_{i})$ is
the PPF of the sub-tree coming into the site $S_{i}$, and $w(\Gamma )$ is the
local weight of the origin square.

Then we can introduce the ratios%

\begin{equation}
x_{m}=\frac{Z_{m}(+)}{Z_{m}(+)+Z_{m}(-)},\text{ }y_{m}=\frac{Z_{m}(-)}%
{Z_{m}(+)+Z_{m}(-)}. \label{Ratios}%
\end{equation}

These two are the ratios of PPFs on the level $m$; they indicate, although not
exactly to be, the probability that the site $S_{m}$ on level $m$ is occupied
by a plus spin ($x_{m}$) or a minus spin ($y_{m}$). Therefore, these two
ratios, which we call the solutions of the lattice, are critical in our model
and all the thermodynamic calculations are based on these solutions.

Define a compact notation for convenience
\begin{equation}
z_{m}(S_{m})=\left\{
\begin{array}
[c]{c}%
x_{m}\text{ if }S_{m}=+1\\
y_{m}\text{ if }S_{m}=-1
\end{array}
\right.  , \label{General Ratio}%
\end{equation}

by introducing%
\[
B_{m}=Z_{m}(+)+Z_{m}(-),
\]
we then have $Z_{m}(+)=B_{m}x_{m}$ and $Z_{m}(-)=B_{m}y_{m}$. Substituting
these two equations into Eq.\ref{PPF+_Recursion} and \ref{PPF-_Recursion} gives%

\[
z_{m}(S_{m})=\frac{B_{m+1}^{2}B_{m+2}}{B_{m}}\underset{\Gamma}{\sum}%
[\underset{\{m+1\}}{\overset{2}{%
{\displaystyle\prod}
}}z_{m+1}(S_{m+1})]z_{m+2}(S_{m+2})w(\Gamma),
\]

where the sum is over $\Gamma=1,2,3,\ldots,8$ for $S_{m}=+1$, and over
$\Gamma=9,10,11,\ldots,16$ for $S_{m}=-1$. It is clear that the solution
(ratios) on one level is a function of the solutions on higher levels in a
recursive fashion.

For convenience, we introduce the polynomials:%

\begin{align}
Q_{m+}(x_{m+1},x_{m+2})  &  =\underset{\Gamma=1}{\overset{8}{\sum}}%
[\underset{\{m+1\}}{\overset{2}{%
{\displaystyle\prod}
}}z_{m+1}(S_{m+1})]z_{m+2}(S_{m+2})w(\Gamma),\label{polynomial Q}\\
Q_{m-}(y_{m+1},y_{m+2})  &  =\underset{\Gamma=9}{\overset{8}{\sum}}%
[\underset{\{m+1\}}{\overset{2}{%
{\displaystyle\prod}
}}z_{m+1}(S_{m+1})]z_{m+2}(S_{m+2})w(\Gamma),\\
Q_{m}(x_{m+1},x_{m+2})  &  =Q_{m+}(x_{m+1},x_{m+2})+Q_{m-}(y_{m+1}%
,y_{m+2})=\frac{B_{m}}{B_{m+1}^{2}B_{m+2}}.
\end{align}

Then the ratio $x_{m}$ is just as function of the ratios on higher level
$x_{m+1}$ and $x_{m+2}$:%

\begin{equation}
x_{m}=\frac{Q_{+}(x_{m+1},x_{m+2})}{Q(x_{m+1},x_{m+2})}.
\label{Ratios_Polynomial_x}%
\end{equation}

\subsubsection{Fix-point solution}

From the recursive relation shown in Eq.\ref{Ratios_Polynomial_x}, we may
expect a repeating solution, which has the form:

$x1,x2,\ldots,x_{v}(x_{v}\geq1)$ and $x_{k}=x_{v+k}$ for some value of
$v\geq1$,

so that Eq.\ref{Ratios_Polynomial_x} holds.

Such a solution can be called a $v$-cycle solution that repeats after the
$v$-th application of the recursive relation equation.

For example, for a 1-cycle solution $x$, we simply have the same ratio $x$ on
all the levels%

\begin{equation}
x=\frac{Q_{+}(x,x)}{Q(x,x)}.\label{1-cycle_solution}%
\end{equation}

For a 2-cycle solution $x_{1}$ and $x_{2}$ we have%

\begin{align}
x_{1} &  =\frac{Q_{+}(x_{2},x_{1})}{Q(x_{2},x_{1})},\label{2-cycle_solution}\\
x_{2} &  =\frac{Q_{+}(x_{1},x_{2})}{Q(x_{1},x_{2})}.\nonumber
\end{align}

That is, the two solutions alternatively appear level by level.

\begin{figure}
[ptb]
\begin{center}
\includegraphics[width=0.8\textwidth]
{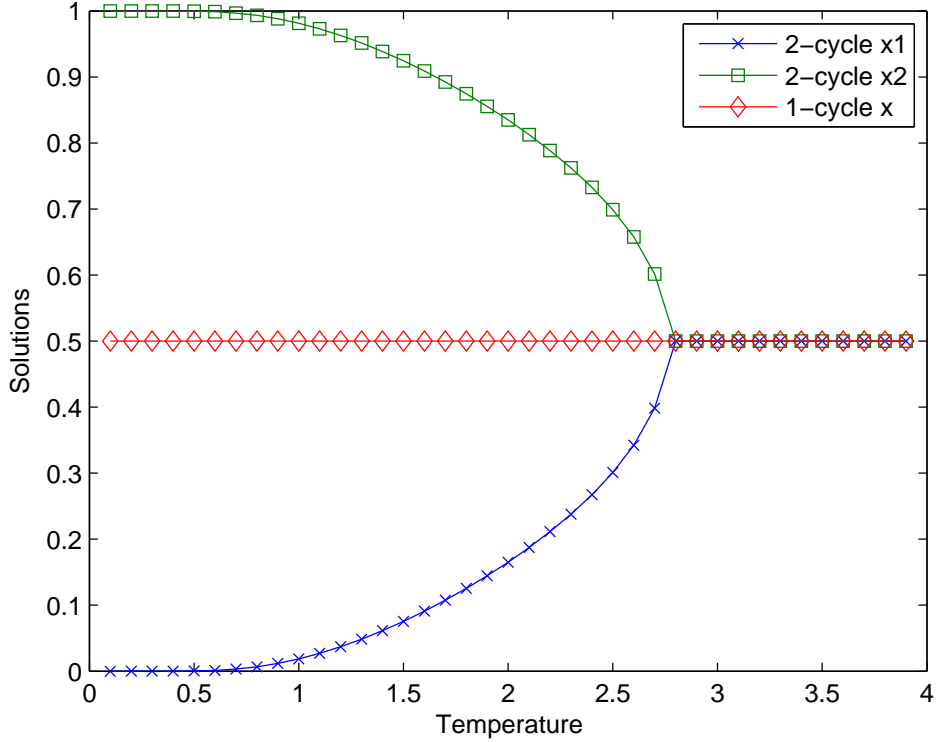}%
\caption{The 1\&2-cycle solutions of the Husimi lattice with $J=-1$ and other
parameters as $0.$}%
\label{fig10}%
\end{center}
\end{figure}

In this way, for a particular recursive relation, we can start from a set of
initial seeds $x_{1}$ and $x_{2}$ to calculate the solutions on lower levels
until we reach the recursively repeating solutions, which is called fix-point
solution. Usually many initial seeds are tried to obtain all the possible
solutions for a particular recursive relation. According to our experience, in
the lattices discussed in this work, solutions with $v\geq3$ were never
obtained, while solutions with $v=1$\ and $v=2$\ ($1$-cycle and $2$-cycle
solutions) are almost always available. Based on the property of $x$, which
determines the probability that one site is occupied by the plus spin, a cycle
solutions with $v\geq3$\ is hard to imagine. An example 1\&2-cycle solutions
of the Husimi lattice, with $J=-1$ and other parameters as $0$ on a wide
temperature range is shown in Fig. \ref{fig10}.%

In Fig. \ref{fig10}, at high temperature both $1$-cycle and $2$-cycle
solutions are $0.5$ and every site in the lattice has a $50\%$ probability to
be occupied by a + spin. This obviously represents a disordered state. At low
temperature, the $1$-cycle solution is still $0.5$, while the $2$-cycle
solution occurs a transition and gives two branches, one of which goes to $1$
and the other goes to $0$ with temperature decrease. This indicates, for
$2$-cycle solution, that two neighbor sites will prefer to have different spin
states. In the region close to zero temperature, if one site has $100\%$
($x=1$) probability to be occupied by $+$ spin, then its neighbor sites will
have $0\%$ ($x=0$) probability to be occupied by $+$ spin (i.e. $100\%$
probability to be occupied by $-$ spin). Therefore, at zero temperature we
will have a plus and minus spins alternative arrangement on the lattice.
Recalling the anti-ferromagnetic model we employed, this $2$-cycle arrangement
has the lowest energy (the most stable state) and corresponds to the ordered
state (crystal). While the $1$-cycle solution at the same temperature refers
to the metastable state, that is, it is still a stable solution however with a
higher energy. The temperature where the $2$-cycle solution appears is called
the critical temperature, or the melting temperature $T_{\text{m}}$. At this
temperature the amorphous state turns to be the ordered state
(crystallization) or it can continue as a metastable state, the analog of
three states here with liquid, crystal and supercooled liquid implies that it
is the melting transition. The thermodynamic details will be discussed in
section III.

\subsection{Recursive calculations on the surface lattice}

\subsubsection{The calculation in the bulk}

For an infinite Husimi lattice, the calculation to get the fix-point solution
is straightforward [25, 37], we can simply start with two artificial initial
seeds and recursively calculate the solutions by Eq.
\ref{Ratios_Polynomial_x} many times until we reach the fix-point solution.
While for the SRL we developed in this paper, the bulk tree is finite and
confined within surface trees, we need to carefully monitor the calculation on
each specific site.

\begin{figure}
[ptb]
\begin{center}
\includegraphics[width=0.8\textwidth]
{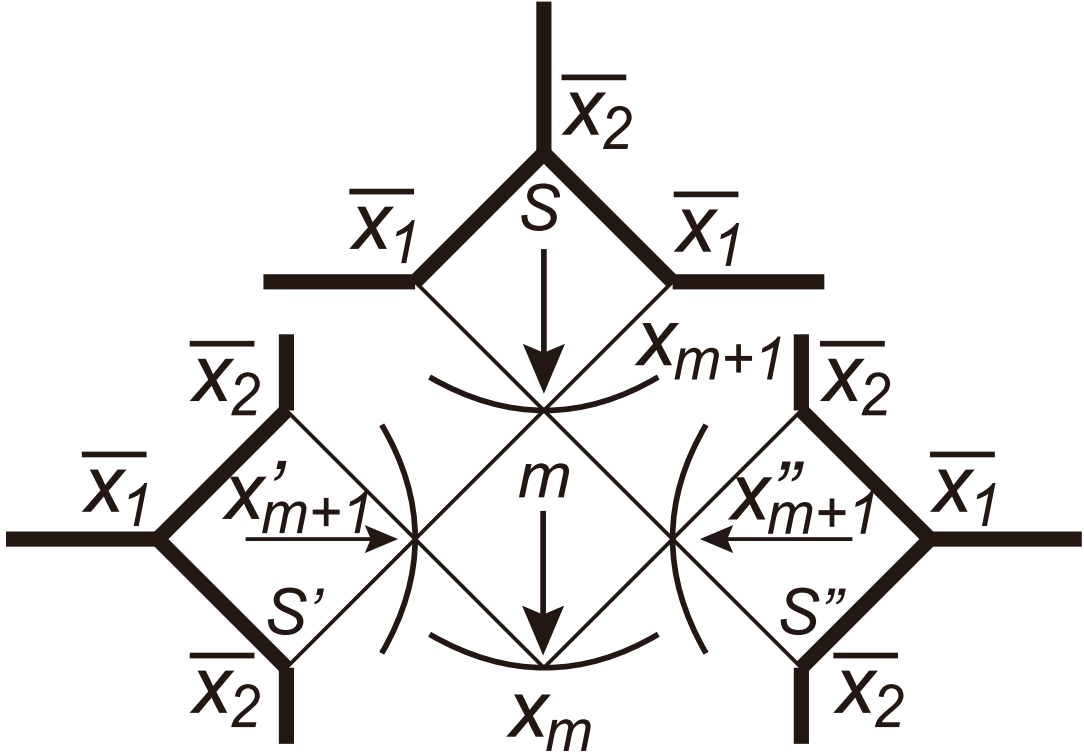}%
\caption{The initial seeds calculation scheme starting from the surface in
SRL}%
\label{fig11}%
\end{center}
\end{figure}

Let us take a SRL with thickness of $2m+3$, we label the surface square as
$S$, then the square next to $S$ is labeled as the $m$-th level, then origin
square is labeled as $0$. Firstly we start from two initial guesses
$\overline{x_{1}}$ $\overline{x_{2}}$ on the surface sites as shown in Fig.
\ref{fig11}. The calculation through the surface square $S$ provides the
solution $x_{m+1}$ on the top site of level $m$-th square. On the other two
surface squares labeled as $S\prime$ and $S"$\ we use the same initial guess
seeds again, however rotate their positions by putting $\overline{x_{1}}$ on
the top site and $\overline{x_{2}}$ on two side sites, then the calculation
based on $S\prime$ or $S"$\ will give us the solution $x\prime_{m+1}$ and
$x"_{m+1}$ on the side sites of level $m$-th square. This is because for the
anti-ferromagnetic Ising model and the properties of $2$-cycle solution
introduced above, we expect a $2$-cycle style solution on our lattice (and the
$1$-cycle solution is just a special case when the two $2$-cycle solutions are
the same), thus the rotation avoids us to get the same results on neighbor
sites, which makes the $2$-cycle solution impossible.%

\begin{figure}
[ptb]
\begin{center}
\includegraphics[width=0.8\textwidth]
{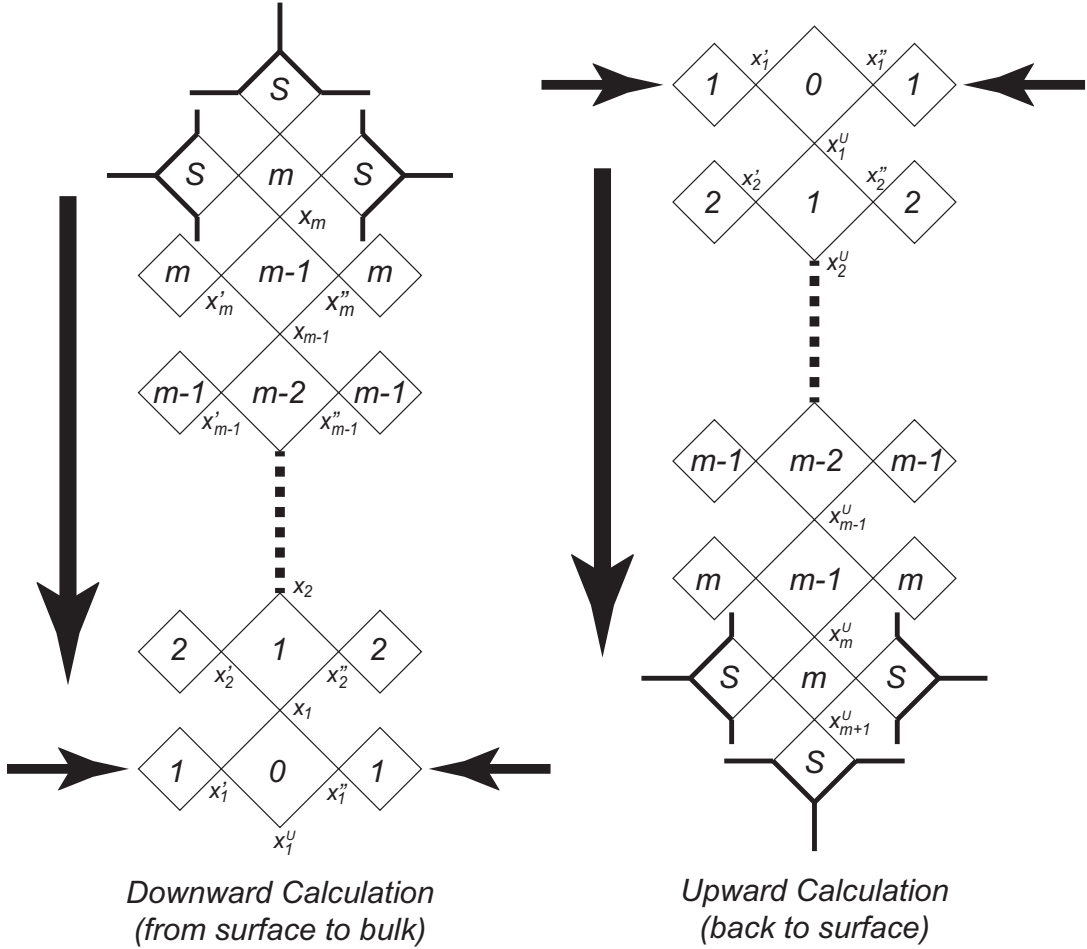}%
\caption{The calculation direction though the bulk tree: (a) start from
surface to the bulk origin, (b) start from the bulk origin and go back to
surface.}%
\label{fig12}%
\end{center}
\end{figure}

Consequently, with the solutions $x_{m+2}$ and $x_{m+1}$ we obtain $x_{m}$ on
the top site of square level $m-1$, and on the other branch the rotation of
$x_{m+2}$ and $x_{m+1}$ will give us $x_{m-1}$, then we can continue the
calculation to the origin square $0$ as shown in Fig. \ref{fig12}(a).

Once we reach the solution $x_{1}$ and $x\prime_{1}$ at the origin square, we
have the states inside the bulk if the solution is the fix-point one. This
part of calculation is called downward calculation, which is from the surface
to the bulk origin. However, this set of bulk solutions is from the initial
guesses on the surface; it may not be the solution of the real state. We still
have to find the fix-point solution on the surface, and the bulk calculation
from the surface fix-point solution is then the final result we are looking
for. To find the fix-point on the surface, we start from the bulk origin and
trace back to the surface (upward calculation), which provides a bulk
contribution \textit{T}$_{\text{B}}$ to the recursive calculation on the surface.%

The upward calculation is shown in Fig. \ref{fig12}(b). At the origin square,
from the contributions of three identical trees and the local weight of the
origin square, we can calculate the solution on the lower site, which is
labeled as $x_{1}^{\text{U}}=f(x1,x\prime1,x"1,w(\Gamma ))$. (Index U stands
for 'upward'). Then from $x_{1}^{\text{U}}$, $x\prime2$ and $x"2$ we can get
$x_{2}^{\text{U}}$ and so on. The upward calculation will provide the solution
$x_{m+1}^{\text{U}}$ at the base site of the surface square unit. This
solution will be used as $x_{\text{B}}$ representing the bulk effect to
surface in the surface calculation.

\subsubsection{The calculation along the surface}

The situation on surface is more complicated than in the bulk. From the
previous introduction on downward and upward calculation, we can see a hint
that depending on the direction of calculation, the solutions might be
different on the same site, for example the $x_{m}$ and $x_{m}^{\text{U}}$ are
different even they are on symmetric sites to the origin. This direction issue
does not affect us to explore the solutions describing the bulk, however it is
critical in surface calculation. Therefore, we firstly classify two directions
on the surface with specific labeling, as shown in Fig. \ref{fig13}:%

\begin{figure}
[ptb]
\begin{center}
\includegraphics[width=0.8\textwidth]
{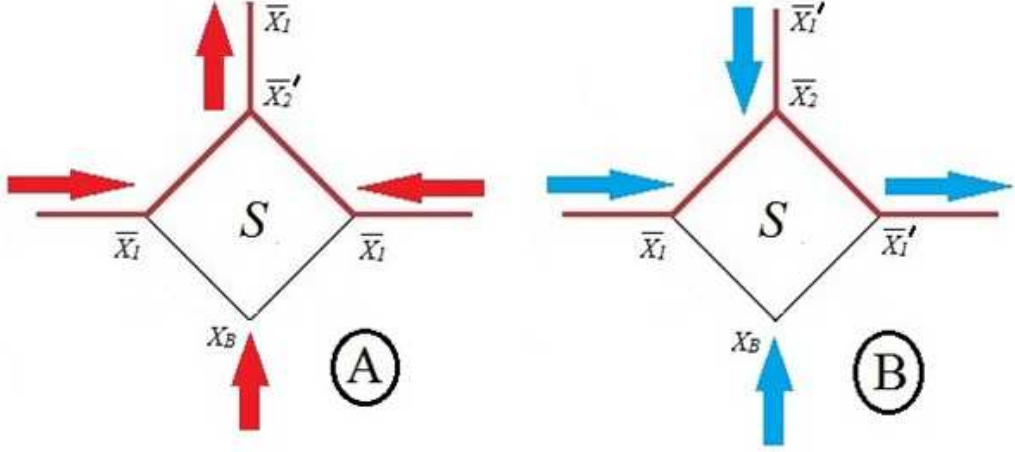}%
\caption{Two calculation directions on the surface of SRL.}%
\label{fig13}%
\end{center}
\end{figure}

\begin{figure}
[ptb]
\begin{center}
\includegraphics[width=0.8\textwidth]
{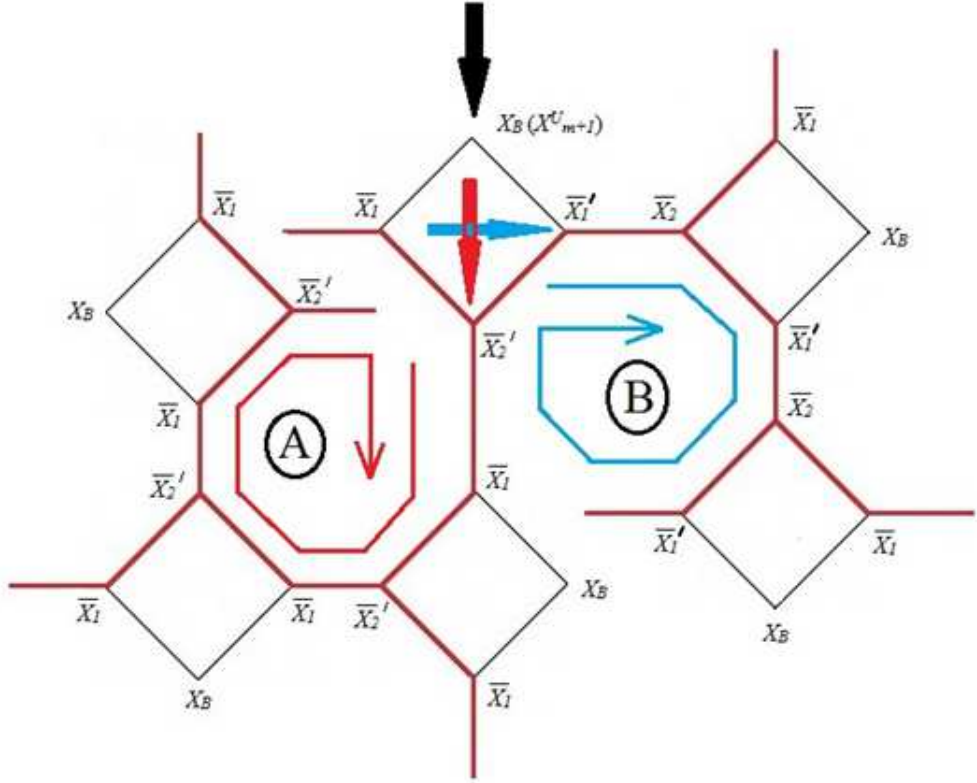}%
\caption{Surface calculation scheme along the surface of SRL.}%
\label{fig14}%
\end{center}
\end{figure}

In Fig. \ref{fig13}, the labeling on the surface is in this way: the solution
on the base site is $x_{\text{B}}$, which is the $x_{m+1}^{\text{U}}$ from
upward calculation; the solution on the sides of the square is label as
$\overline{x_{1}}$, and on the top is $\overline{x_{2}}$; the prime marks the
direction that goes out from the square, while the label without prime is the
direction going into the square. The scheme A shows a \textquotedblleft
side-to-top\textquotedblright\ direction: the solution going out from the top
site, $\overline{x_{2}}^{\prime}$, is calculated from the solution going into
the side site $\overline{x_{1}}$ and the base site solution $x_{\text{B}}$.
Then from $\overline{x_{2}}^{\prime}$ we go through a single bond to get the
next $\overline{x_{1}}$. The scheme B shows a \textquotedblleft
side\&top-to-side\textquotedblright\ direction: the solution going out from
the side site, $\overline{x_{1}}^{\prime}$, is calculated from the solution
going into the side site $\overline{x_{1}}$, the solution going into the top
site $\overline{x_{2}}$, and the base site solution $x_{\text{B}}$. Again from
$\overline{x_{1}}^{\prime}$ we then go through a single bond to get the next
$\overline{x_{2}}$. Only after we have done the calculations in both
directions, we can get the surface solutions pair, the solution going into the
top site $\overline{x_{2}}$, and the solution going into the side site
$\overline{x_{1}}$, to do the subsequent bulk calculation. (Note that
initially this pair is a guess we made to do the first iteration's bulk
calculation). The surface calculation scheme is shown in Fig. \ref{fig14}.

The upward bulk calculation provides us the bulk tree contribution and the
solution $x_{\text{B}}$ on the base site of surface square. This solution can
be taken as a constant in the surface recursive calculation. In Fig.
\ref{fig14}, we start from the $x_{\text{B}}$ and the initial seed
$\overline{x_{1}}$ to do two calculations as mentioned. The recursive
calculation in process A will eventually give us a fixed $\overline{x_{1}}$,
while the process B will give us a fixed $\overline{x_{2}}$. This new fixed
set of $\overline{x_{1}}$ and $\overline{x_{2}}$ is then the seeds we are
going to use for the next iteration's bulk calculation. Note one difference in
process A and B is that in process A we only need $\overline{x_{1}}$ for
calculation, which is updated step by step, however in process B after each
step we will only have a updated $\overline{x_{1}}^{\prime}$, while
$\overline{x_{1}}$ is a constant in calculation. Thus, in practice we always
do the process A first to get a fixed $\overline{x_{1}}$, then use this
$\overline{x_{1}}$ as a constant in the calculation of process B.%

The calculation on the square unit, for example, $\overline{x_{1}}^{\prime
}=f(\overline{x_{1}},\overline{x_{2}},x_{\text{B}})$ in process B, is the same
as the bulk calculation. The calculation through the single surface bond, for
example, $\overline{x_{2}}=f(\overline{x_{1}}^{\prime})$ in process B, is also
similar with the square. The difference is that the PPF only has two terms
since there are $4$ configurations of a single bond structure:%

\begin{align*}
Z_{m}(+)  &  =\underset{\Gamma=1}{\overset{2}{\sum}}Z_{m+1}(S_{m+1}%
)w(\Gamma),\\
Z_{m}(-)  &  =\underset{\Gamma=3}{\overset{2}{\sum}}Z_{m+1}(S_{m+1})w(\Gamma),
\end{align*}

where the sum is over $\Gamma=1$ and $2$ for $S_{m}=+1$ , and over $\Gamma=3$
and $4$ for $S_{m}=-1$ . Then we have the polynomials:%

\begin{align*}
Q_{m+}(x_{m+1})  &  =\underset{\Gamma=1}{\overset{2}{\sum}}z_{m+1}%
(S_{m+1})w(\Gamma),\\
Q_{m-}(y_{m+1})  &  =\underset{\Gamma=3}{\overset{2}{\sum}}z_{m+1}%
(S_{m+1})w(\Gamma),\\
Q_{m}(x_{m+1})  &  =Q_{m+}(x_{m+1})+Q_{m-}(y_{m+1}).
\end{align*}

The ratio is the function of the ratio on higher level:%

\[
x_{m}=\frac{Q_{m+}(x_{m+1})}{Q_{m}(x_{m+1})}.
\]

\begin{figure}
[ptb]
\begin{center}
\includegraphics[width=0.8\textwidth]
{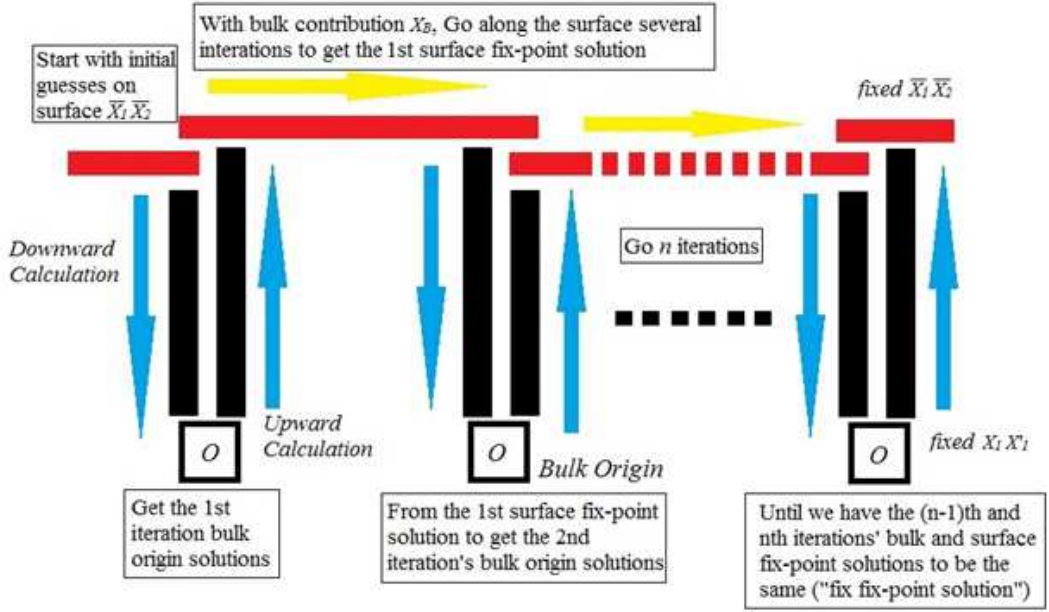}%
\caption{The calculation process scheme to reach the fix-point solutions in
SRL. The vertical process is calculations through bulk tree and the horizontal
process is along the surface.}%
\label{fig15}%
\end{center}
\end{figure}

The calculation scheme to reach the fix-point solutions both on the surface
and in the bulk is shown in Fig. \ref{fig15}. We do an embedded recursive
calculations to get the final solutions on the surface and in the bulk. The
$n$ iterations give us recursively updated solutions in the bulk until we find
the fix-point solution, and a recursive calculation along the surface is
embedded in one iteration to take the effect of updated bulk contributions
each time. This process is to make sure that we counter the mutual effects
from surface to bulk and from bulk to surface. Another necessity of this
complicated process is that the bulk tree has a finite size thus the fix-point
solution is not guaranteed to be reached during one downward calculation. For
the structure with thickness $\geq19$, we can always obtain the fix-point
solution at bulk origin, although on the first several layers close to surface
we can only have numerical calculations instead of exact solutions. (This
`numerical depth' depends on the energy parameters setting and it shows always
being smaller than $19$ according to our experience.)%

\section{THERMODYNAMIC CALCULATION}

\subsection{General free energy calculation method: the Gujrati trick}

Since our lattices are infinitely large, it makes no sense to calculate the
partition function or free energy for the entire lattice. However an exact
treatment called Gujrati trick has been well developed to deal with the
thermodynamic calculation on recursive lattices in our group's previous work
\cite{29,37,49,50}. By
this technique we can approach the thermal properties in a local area (per
site) by the partial partition functions $Z(S)$ and solutions $x$ we discussed
in the previous section.

The Helmholtz free energy is a function of the temperature and partition
function $Z$:%

\[
F=-T\log Z
\]
%

\begin{figure}
[ptb]
\begin{center}
\includegraphics[width=0.4\textwidth]
{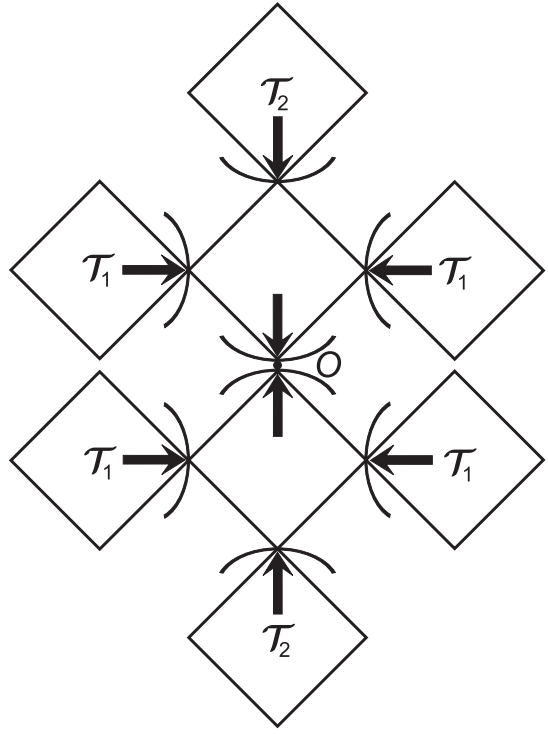}%
\caption{The sub-tree contributions to the origin point.}%
\label{fig16}%
\end{center}
\end{figure}

Here we take a regular Husimi lattice as an example to introduce the thermal
calculation derived by figuring out the partition functions. Recall Eq.
\ref{PF_of_PPFs}, since $Z_{0}$ counts the contribution of the whole system,
it must be the sum of contributions of the six sub-trees on level $1$ and $2$,
and the weight of the two local square units on the origin. If we cut off the
two sub-branches on level $2$ and joint them together, we will have an
identical but smaller lattice. Similarly the partition function of this
smaller lattice is:%

\[
Z_{2}=Z_{2}^{2}(+)\exp(\beta H)+Z_{2}^{2}(-)\exp(-\beta H).
\]

We also cut off the 4 sub-branches on level 1 and form two smaller lattices
and their partition function will be%

\[
Z_{1}=Z_{1}^{2}(+)\exp(\beta H)+Z_{1}^{2}(-)\exp(-\beta H).
\]

As an extensive quantity the free energy of the whole system is the sum of the
free energies of these three smaller lattices and the two local squares:%

\[
F(Z_{0})=2F(Z_{1})+F(Z_{2})+F(Z_{\text{local}})
\]

This yields%

\[
F_{\text{local}}=-T\log(\frac{Z_{0}}{Z_{1}^{2}Z_{2}}).
\]

as the free energy of the two local sqaures, i.e. four sites (three paris of
half-sites and the origin site).

The free energy per site is:%

\[
F=-\frac{1}{4}T\log(\frac{Z_{0}}{Z_{1}^{2}Z_{2}}).
\]

By substituting%

\[
Z_{m}(+)=B_{m}x_{m}%
\]

and%

\[
Z_{m}(-)=B_{m}(1-x_{m}),
\]

we have%

\begin{equation}
F=-\frac{1}{4}T\log(\frac{B_{0}^{2}[x_{0}^{2}e^{\beta H}+(1-x_{0}%
)^{2}e^{-\beta H}]}{B_{1}^{4}[x_{1}^{2}e^{\beta H}+(1-x_{1})^{2}e^{-\beta
H}]^{2}B_{2}^{2}[x_{2}^{2}e^{\beta H}+(1-x_{2})^{2}e^{-\beta H}]})
\label{FreeEnergyPolynomial/site}%
\end{equation}

recall that for either 1-cycle or 2-cycle fix-point solutions we have
$x_{0}=x_{2}$, and%

\[
Q_{0}=\frac{B_{0}}{B_{1}^{2}B_{2}}.
\]

It follows%

\[
F=-\frac{1}{2}T\log(Q_{0}\frac{1}{x_{1}^{2}e^{\beta H}+(1-x_{1})^{2}e^{-\beta
H}}).
\]

With the calculation of $Q_{0}$ and fix-point solution $x_{0}$\ and $x_{1}$ we
discussed in the previous section, the free energy can be easily achieved.

For a recursive structure with an origin point, we can always do this
\textquotedblleft cut and rearrange\textquotedblright\ trick to obtain the
partition function and free energy per site around the origin point.

The entropy is the first derivative of the free energy with respect to the temperature:%

\begin{equation}
S=-\partial F/\partial T.\label{entropy}%
\end{equation}

With the free energy and entropy we have the energy per site (energy density) as%

\begin{equation}
E=F+TS.\label{energy}%
\end{equation}

A typical thermodynamic behavior of a Husimi square lattice with $J=-1$ and
all other parameters to be $0$ is shown in Fig. \ref{fig17}:%

\begin{figure}
[ptb]
\begin{center}
\includegraphics[width=0.8\textwidth]
{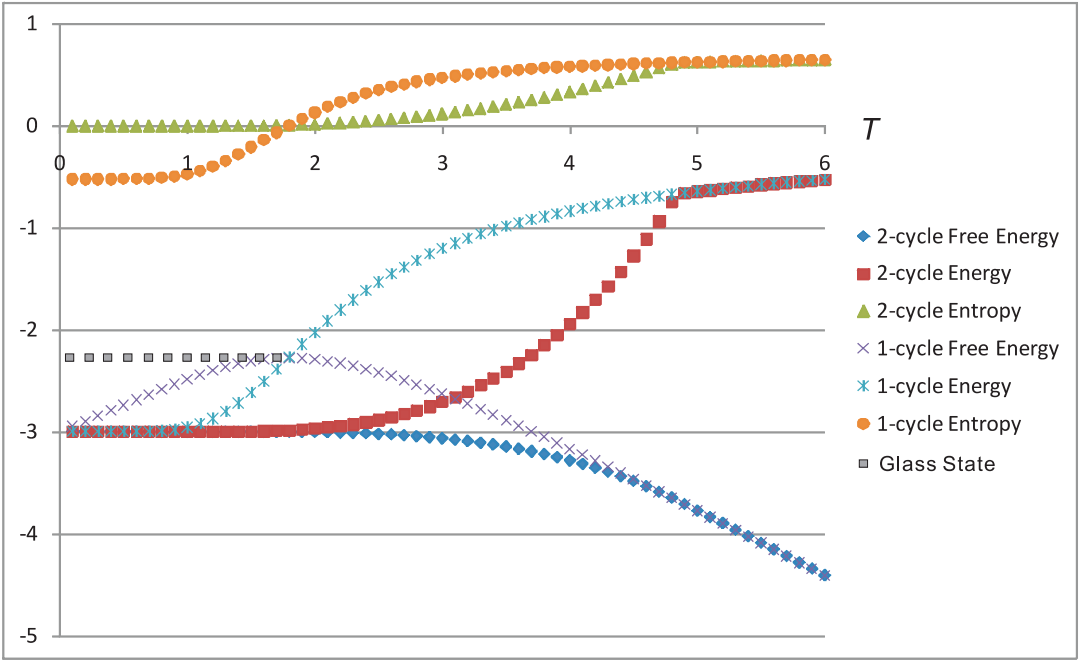}%
\caption{The thermodynamic behavior of a Husimi square lattice with $J=-1$ and
all other parameters to be $0.$}%
\label{fig17}%
\end{center}
\end{figure}

In Fig. \ref{fig17}, the free energy of two solutions are the same at high
temperature. At the melting temperature $T_{\text{m}}$ the 1-cycle's free
energy differs from 2-cycle's and is less stable. The continuous entropy at
$T_{\text{m}}$ of 1-cycle implies the supercooled liquid state, while the
entropy decrease of 2-cycle state implies the crystallization. Although here
the entropy decrease is not a discontinuous jump as the typical behavior of
first order melting transition, we may still treat it as the phases-coexisting
point since the 2-cycle state is the most stable state here.

The thermodynamics of 1 and 2-cycle solutions agree with our
anti-ferromagnetic model. The anti-ferromagnetic Ising model prefer to
anti-aligned neighbor spins, therefore the 2-cycle state should have the
lowest energy, i.e. the crystal, while the 1-cycle solution represents the
metastable state with higher energy. With the continuing decrease of
temperature we can observe that at $T=1.1$ the entropy of 1-cycle state
(supercooled liquid) quickly decreases to zero and becomes negative. This is
the Kauzmann's paradox and the ideal glass transition. The negative entropy is
unphysical thus the metastable state must undergo a transition to be the glass
state at $T_{\text{K}}$, the ideal glass transition temperature. The free
energy of the glass state below glass transition temperature is indicated by
the last legend in Fig. \ref{fig17}. Note that this branch is not provided by
the calculation, instead it is a theoretical expectation. We can modify our
calculation to achieve a stable solution of this glass state branch, but since
this part is not our interest we have not done that calculation in this paper.

\subsection{The Gujrati trick applied on the SRL}

We are following the Gujrati trick to solve the free energy and consequent
thermal functions of the SRL. Although the basic principle is the same, the
asymmetrical structure of the surface lattice requires further complex tricks
to do the calculation. If we take a random site on the surface as the origin,
to do the \textquotedblleft cut and re-joint\textquotedblright\ trick we need
to select a local area around the origin, and find matching sub-trees
contributing to the local area. Here the matchings are more specific than they
are in a homogeneous bulk lattice. For example in the Husimi lattice discussed
in previous section, all the sites are identical, thus if two sites have the
same solution $x$ on them, the sub-trees cut off from them can be joined
together to make a new lattice. However, the sites on the surface of SRL have
three different situations: on the top of a surface square where the
sub-tree's contribution going out, on the side of a surface square where the
sub-tree's contribution coming into, and on the bottom of a surface square
where the bulk tree's contribution coming into. The matchings must be done on
sites with exactly the same situations to make an identical but small lattice,
but the asymmetrical structure of SRL makes the cutting and matching
impossible. Wherever we select the origin local area and cut the sub-trees,
there will always be two sub-trees left and cannot be matched with each another.%

\begin{figure}
[ptb]
\begin{center}
\includegraphics[width=0.8\textwidth]
{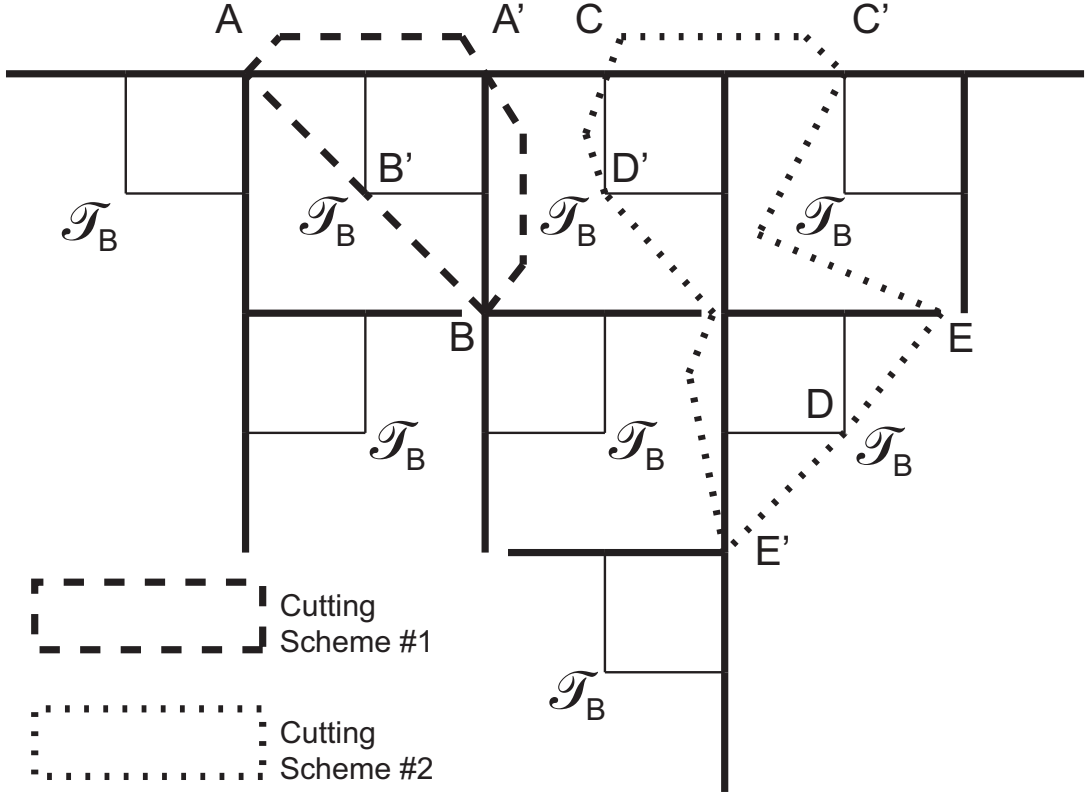}%
\caption{Two example failure cuttings in SRL.}%
\label{fig18}%
\end{center}
\end{figure}

Fig. \ref{fig18} shows two failure cutting schemes, with \textquotedblleft1
square and 2 single bonds\textquotedblright\ area and \textquotedblleft2
squares and 4 single bonds\textquotedblright\ area respectively. In the
cutting scheme \#1, by removing the local area surrounded by the discontinuous
curves, we have four sub-tree at the point A, A', B and B'. The two sub-trees
on A and A' can be linked together to form a smaller lattice with exactly the
same structure. While the sub-trees on B and B' do not match with each other,
since they are a bulk sub-tree on B and a surface sub-tree contributing the
top site of a surface square.

In the cutting scheme \#2, the local area of \textquotedblleft2 squares and 4
single bonds\textquotedblright\ is bounded by the solid curves. The sub-trees
on C - C' and D - D' can form smaller lattices. Although the new formation of
D - D' is not identical to the entire lattice, we can still count its
partition function and handle the ratio calculation. However the same
difficulty, as we encountered in scheme \#1, presents on the sites E and E'.
The E site is at the side of a surface square while the E' site is on the top.

\begin{figure}
[ptb]
\begin{center}
\includegraphics[width=0.4\textwidth]
{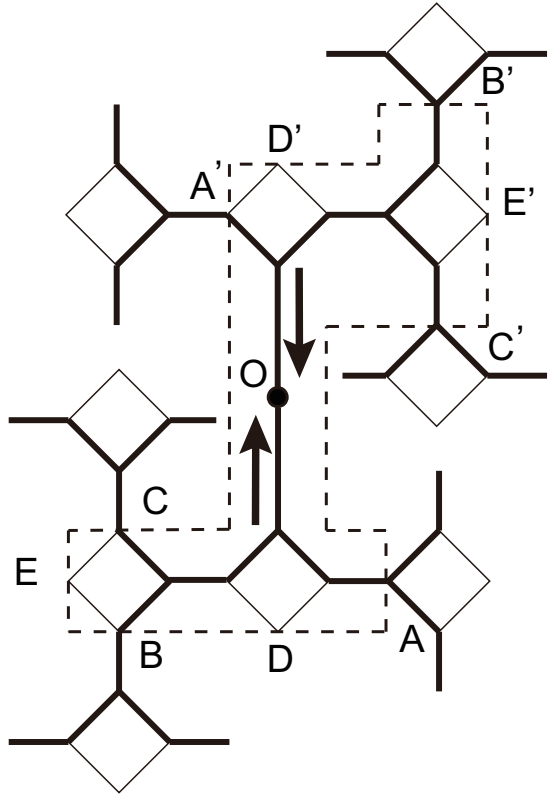}%
\caption{The modification on the surface of lattice and the cutting area
around the origin.}%
\label{fig19}%
\end{center}
\end{figure}

Therefore, either we choose odd or even numbers of basic units as the local
area around origin, the Gujrati trick cannot be done. We have to somehow
modify the origin structure to avoid the problem of asymmetry. As shown in
Fig. \ref{fig13}, the single bond and square unit have a \textquotedblleft
head-to-side\textquotedblright\ connection, that is, following the calculation
direction, the sub-tree coming from a single bond is always linked to the side
site of next level's square, or vice versa. This arrangement is designed to
make the structure uniform on the surface, but it also causes the asymmetry.
It is necessary to make one \textquotedblleft head-to-head\textquotedblright%
\ connection as the origin of the surface, as shown in Fig. \ref{fig19}.%

In this way, we have two infinite and identical surface trees connected in a
\textquotedblleft head-to-head\textquotedblright\ style, this single bond is
then the origin of our entire lattice, and except for this bond, all other
surface single bonds are in \textquotedblleft head-to-side\textquotedblright%
\ connection. Starting from the origin, the closest squares can be labeled as
level $1$; the second closest squares are on level $2$ and so on. As shown in
Fig. \ref{fig19}, by the selection of four squares and six single bonds as the
local area, we can form one identical lattice on level $1$ at the sites A -
A', two identical lattices on level two at the pairs B - B' and C - C'. The
sub-trees cut at D, D', E and E' are identical bulk trees so we can pair them
to form two bulk lattices. Now we can easily derive the free energy of the
local area to be:%

\begin{equation}
F_{\text{local}}=-T\log(\frac{Z_{0}}{Z_{1}Z_{2}^{2}Z_{\text{B}}^{2}%
}).\label{FE_local_surface}%
\end{equation}

Unlike the point origin in Husimi lattice, here the origin is a single bond
unit. The total partition function thus has four terms due to the four
possible states of the single bond:%

\begin{equation}
Z_{0}=Z_{0}^{\text{U}}(+)Z_{0}^{\text{L}}(-)\exp(-\beta\overline{J}%
)+Z_{0}^{\text{U}}(-)Z_{0}^{\text{L}}(+)\exp(-\beta\overline{J})+Z_{0}%
^{\text{U}}(+)Z_{0}^{\text{L}}(+)\exp[\beta(\overline{J}+2\overline{H}%
)]+Z_{0}^{\text{U}}(-)Z_{0}^{\text{L}}(-)\exp[\beta(\overline{J}-2\overline
{H})] \label{PF_of_U_and_L_PPFs}%
\end{equation}

\begin{figure}
[ptb]
\begin{center}
\includegraphics[width=0.4\textwidth]
{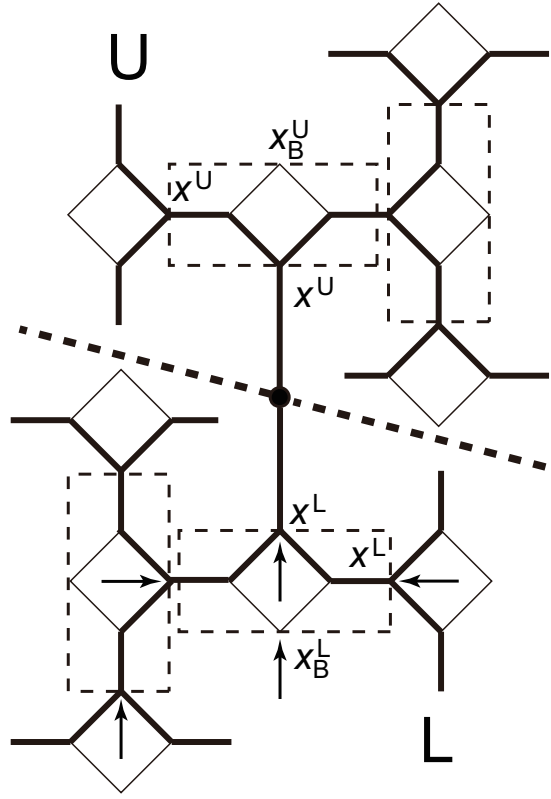}%
\caption{The redefined unit in SRL for the thermodynamic calculations around
the origin.}%
\label{fig20}%
\end{center}
\end{figure}

To specifically track the spins in the origin area and rematch the sub-trees
we define the two sub-trees meeting at the origin bond as \textquotedblleft
upper\textquotedblright\ and \textquotedblleft lower\textquotedblright\ parts.
In equation \ref{PF_of_U_and_L_PPFs}, the superscript U or L is to indicate
the upper or lower part of the lattice. Similar integration can be employed to
form the partition functions of smaller lattice we rearranged in
Eq.\ref{FE_local_surface}. The ratio of partition functions can be
represented by the solutions and polynomials which have been achieved in
Section II. However in Section II, we go along the surface and calculate the
solution on each site and the polynomial ratios step by step, these detailed
solutions are unnecessary here and make it complicated to determine the
Eq.\ref{FE_local_surface}. Hence we redefine the basic unit as one square and
two single bonds linked on its side as shown in Fig. \ref{fig20}.%

By the selection of this rectangle basic unit, in the upper or lower branch we
simply have two units contributing to the sides of next unit, and recursively
the whole branch contributing to the origin bond, as indicated by the arrows
in the lower part in Fig. \ref{fig20}. The contribution of bulk tree can be
treated as a constant. The calculation based on this selection will provide
only one solution as $x^{\text{L}}$ or $x^{\text{U}}$ on the output site,
regardless of it is 1-cycle and 2-cycle solutions on the surface. This single
solution is sufficient for us to determine the Eq.\ref{FE_local_surface}. For
instance, if we look at the smaller lattices rearranged on points A - A' in
Fig. \ref{fig20}, we will only need the solution $x^{\text{L}}$ and
$x^{\text{U}}$ because they are on the output sites A and A'.

Similarly, for either $x^{\text{L}}$ or $x^{\text{U}}$, recall the
Eq.\ref{General Ratio} and $Z_{m}(+)=B_{m}x_{m}$, $Z_{m}(-)=B_{m}y_{m}$ with
$B_{m}=[Z_{m}(+)+Z_{m}(-)]$, then we have%

\[
z_{m}(S_{m})=\frac{B_{m+1}^{2}B_{\text{B}}}{B_{m}}\underset{\Gamma}{%
{\displaystyle\sum}
}[\overset{2}{\underset{\{m+1\}}{%
{\displaystyle\prod}
}}z_{m+1}(S_{m+1})]z_{\text{B}}(S_{\text{B}})\omega(\Gamma),
\]

where the $S_{\text{B}}$ is the index of the site linked to the bulk tree.

With six spins the rectangle unit has $64$ possible configurations thus we can
introduce the polynomials:%

\[
Q_{m+}(x_{m+1})=\underset{\Gamma=1}{\overset{32}{%
{\displaystyle\sum}
}}[\overset{2}{\underset{\{m+1\}}{%
{\displaystyle\prod}
}}z_{m+1}(S_{m+1})]z_{\text{B}}(S_{\text{B}})\omega(\Gamma),
\]

\[
Q_{m-}(y_{m+1})=\underset{\Gamma=33}{\overset{32}{%
{\displaystyle\sum}
}}[\overset{2}{\underset{\{m+1\}}{%
{\displaystyle\prod}
}}z_{m+1}(S_{m+1})]z_{\text{B}}(S_{\text{B}})\omega(\Gamma),
\]

\[
Q_{m}(x_{m+1})=Q_{m+}(x_{m+1})+Q_{m-}(y_{m+1})=\frac{B_{m}}{B_{m+1}%
^{2}B_{\text{B}}}.
\]

Then we can obtain the 1-cycle recursive relation:%

\[
x_{m}=\frac{Q_{m+}(x_{m+1})}{Q_{m}(x_{m+1})}%
\]

And the fix-point solution is the one cycle solution $x^{\text{L}}$ or
$x^{\text{U}}$, Although the actual solution site by site on the surface could
be either 1-cycle or 2-cycle. It is important to clarify that the solution
$x^{\text{L}}$ or $x^{\text{U}}$ are exactly the same to the solutions we
achieved on the surface in Section II. The reason we re-select the basic unit
is to obtain a simple polynomial $Q$ for the calculation in equation (3.13)
and (3.14).

Now we may rewrite Eq.\ref{FE_local_surface} with $\overline{H}=0$%

\begin{equation}
\begin{split}
F_{\text{local}}=-T\log[\frac{Q_{m}^{U}Q_{m}^{L}}{(x^{\text{U}}y^{\text{L}%
}\exp(-\beta\overline{J})+y^{\text{U}}x^{\text{L}}\exp(-\beta\overline
{J})+x^{\text{U}}x^{\text{L}}\exp\beta\overline{J}+y^{\text{U}}y^{\text{L}%
}\exp\beta\overline{J})^{2}}
\\\cdot\frac{1}{(x_{\text{B}}^{\text{U}}x_{\text{B}}^{\text{U}}%
\exp(\beta H)+y_{\text{B}}^{\text{U}}y_{\text{B}}^{\text{U}}\exp(-\beta
H))(x_{\text{B}}^{\text{L}}x_{\text{B}}^{\text{L}}\exp(\beta H)+y_{\text{B}%
}^{\text{L}}y_{\text{B}}^{\text{L}}\exp(-\beta H))}].
\end{split}
\end{equation}

This local area contains $14$ sites, thus the averaged free energy on each
site is%

\[
F=\frac{1}{14}F_{\text{local}}.
\]

\begin{figure}
[ptb]
\begin{center}
\includegraphics[width=0.8\textwidth]
{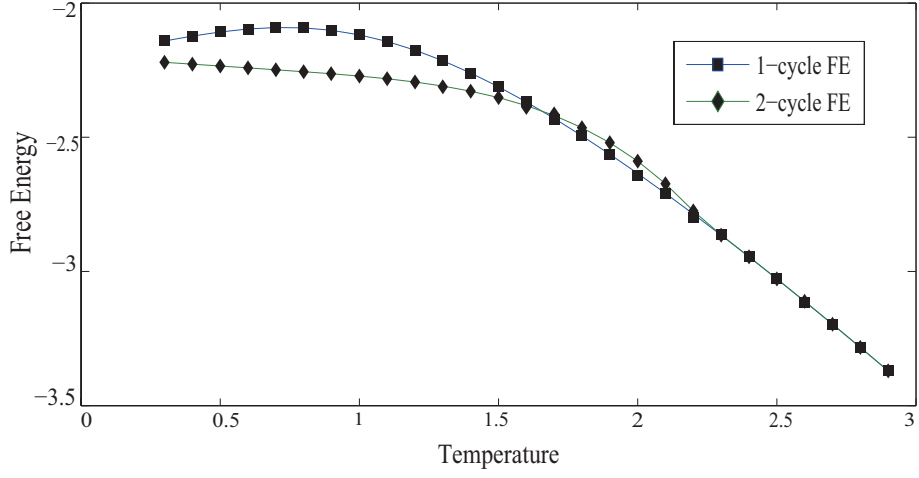}%
\caption{The free energy behaviors of 1-cycle and 2-cycle solution of SRL with
thickness $=19$, $J=-1$, $\overline{J}=-1$ and all other parameters to be
zero}%
\label{fig21}%
\end{center}
\end{figure}

An reference free energy behaviors of 1-cycle and 2-cycle solution of SRL with
thickness =19, neighbor interaction $J=-1$, surface neighbor interaction
$\overline{J}=-1$ and all other parameters to be zero is shown in Fig.
\ref{fig21}%

An interesting phenomenon can be observed from Fig. 3.6. The free energy of
2-cycle solution (crystal state) is higher than the free energy of metastable
state between T = $1.65$ to $2.2$. This is different from the bulk behavior in
Fig. \ref{fig21}. The cross point of 1 and 2-cycle's free energy at lower
temperature can be determined to be the melting transition. Above the melting
temperature, our results indicate that the anti-aligned spins arrangement is
less stable than the $0.5$ solution. This behavior has not been investigated,
yet it is not clear whether it is simply unphysical or it implies a
\textquotedblleft super-heated crystal\textquotedblright\ state.

Below the melting temperature the behavior is similar to the bulk system. The
2-cycle solution represents the crystal state with a lower free energy and
continues to decrease to the free energy of ideal crystal. While the free
energy of 1-cycle solution, the metastable state, will reach a minimum point
than bind back, this unphysical behavior implies the ideal glass transition.

\begin{figure}
[ptb]
\begin{center}
\includegraphics[width=0.8\textwidth]
{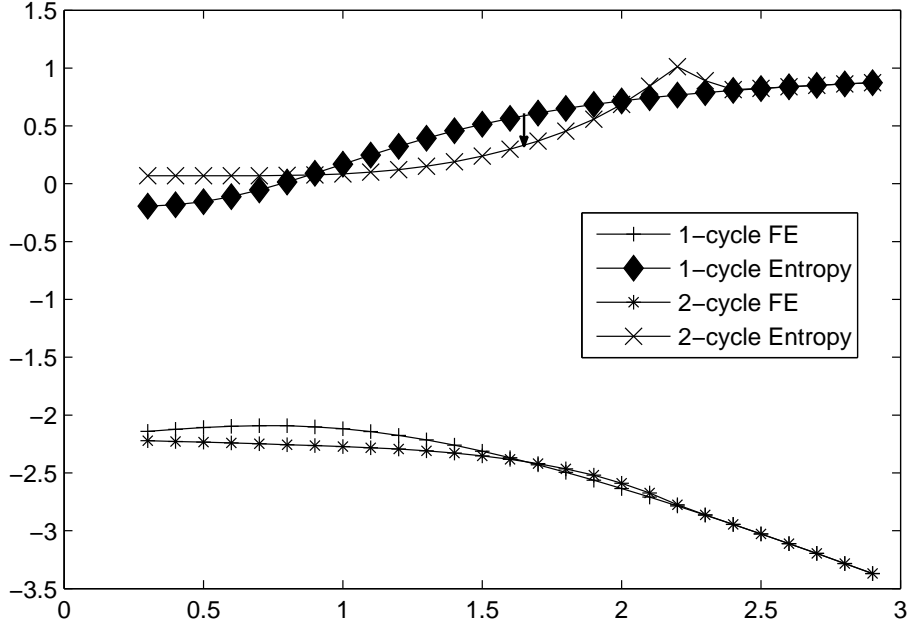}%
\caption{The thermodynamic behaviors of SRL with $J=-1$, $\overline{J}=-1$ and
all other parameters to be zero.}%
\label{fig22}%
\end{center}
\end{figure}

With equations \ref{entropy} and \ref{energy} we can easily calculate the
energy density and entropy from the free energy. The Fig. \ref{fig22} shows
the entropy derived from the free energy in Fig. \ref{fig21}. In Fig.
\ref{fig22} the black arrow on the entropy show the melting transition at the
cross point of free energies ($T=1.65$), where the entropy of 1-cycle solution
will step to 2-cycle solution's entropy as a first order transition. The ideal
glass transition temperature $T_{\text{K}}$ can be clearly observed by the
negative entropy of 1-cycle solution. The detailed discussion on surface
thermodynamics and the surface effect comparing with bulk system will be
presented in the next section.%

\section{RESULTS AND DISCUSSION}

\subsection{Discussion on the solutions}

\begin{figure}
[ptb]
\begin{center}
\includegraphics[width=0.8\textwidth]
{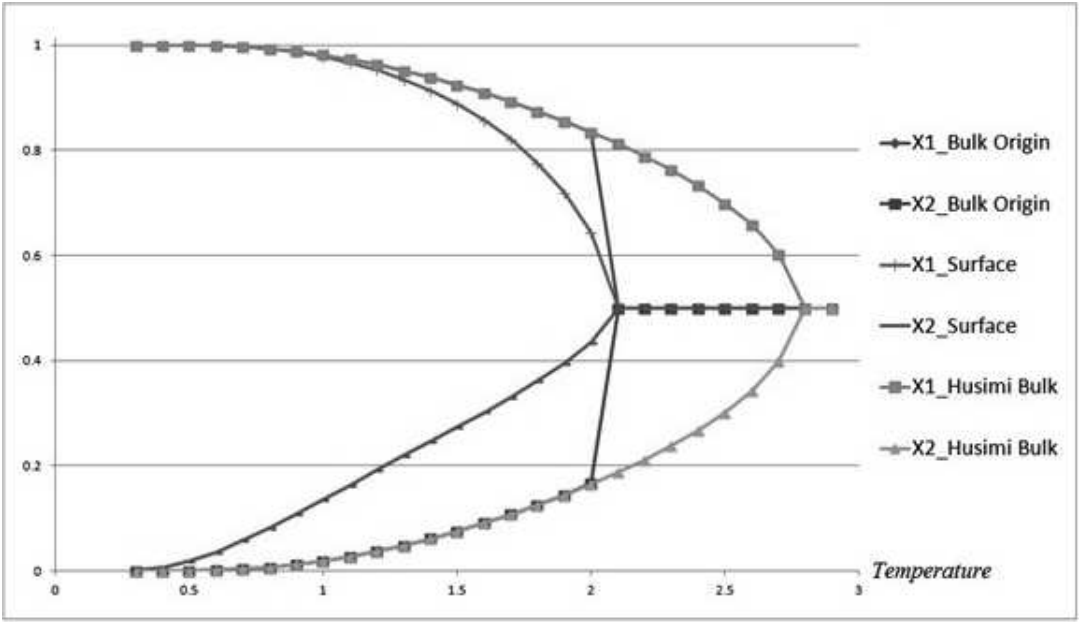}%
\caption{The reference 2-cycle solutions of SRL with J and J =-1, other
parameters as 0, and thickness = 19.}%
\label{fig23}%
\end{center}
\end{figure}

As introduced in previous chapters, the thermodynamic calculations are mainly
based on the solutions $x$, i.e. the ratio of partial partition functions on
SRL. The 1 and 2-cycle solutions of a simple anti-magnetic field Husimi case
is presented in Fig. \ref{fig10}, and we have discussed how the melting
transition, crystal state and metastable state can be indicated from the
solutions. The reference 2-cycle solutions of SRL with $J$ and $\overline
{J}=-1$, other parameters as $0$, and thickness = $19$ are shown in Fig.
\ref{fig23}.%

The solutions of bulk Husimi lattice from Fig. \ref{fig10} are also included
in Fig. \ref{fig23} for comparison. With the temperature increase, the
solutions on surface converge to $0.5$ at lower temperature than the Husimi
bulk solutions. This indicates a lower melting transition temperature on the
surface, which is easy to understand with smaller coordination number and less
interactions on the surface, the spins on the surface are easier to be
anti-aligned with less energy. Also, unlike the symmetrical solutions we
usually have, the 2-cycle solution on the surface is asymmetric. This is
because of the hybrid structure on the surface, which defines two different
circumstances on the surface square unit and single bond unit respectively.
Naturally the spins on the surface square unit has the solution closer to the
bulk solution, while the spins on the single bond unit has a less stable
(closer to $0.5$) solution due to the lower dimension. Depending on the
initial seeds adopted for surface calculation, we may also have the other set
of surface solutions symmetric to the one in \ref{fig23}.

\begin{figure}
[ptb]
\begin{center}
\includegraphics[width=0.8\textwidth]
{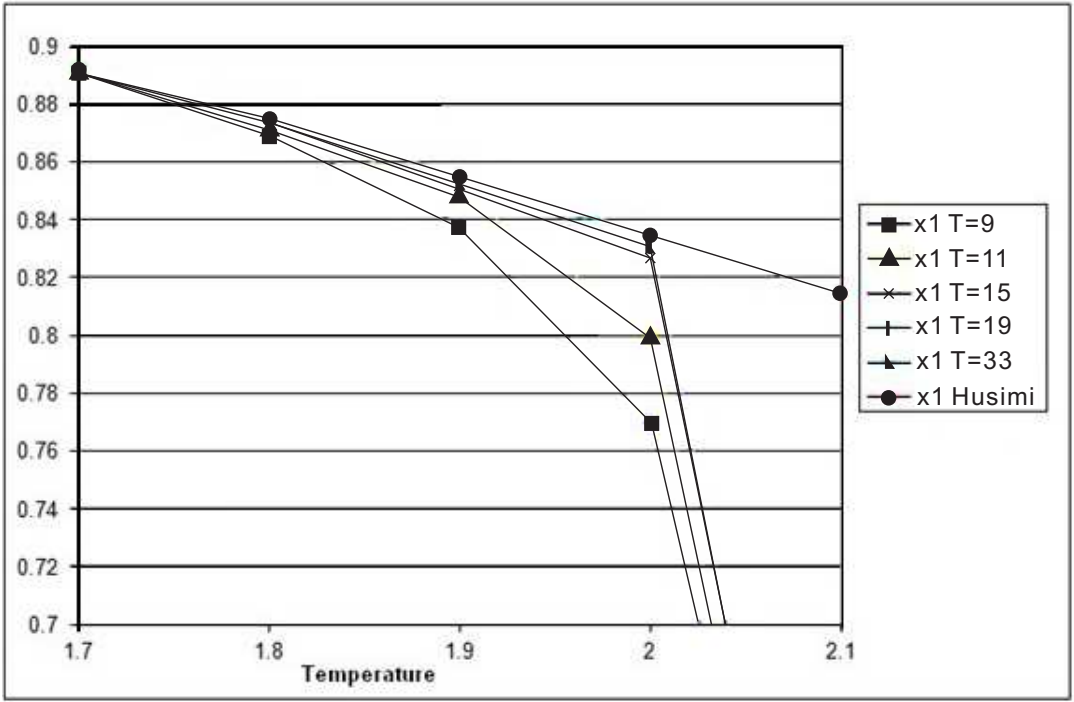}%
\caption{The bulk origin solutions (upper branch) at converging point with
different thickness. The curves of thickness = 19 and 33 almost overlap and
cannot be distinguished in the graph.}%
\label{fig24}%
\end{center}
\end{figure}

The 2-cycle bulk solutions $x_{1}^{\text{BULK}}$ and $x_{2}^{\text{BULK}}$ are
the solutions at the bulk origin (Fig. \ref{fig15}). Since the Husimi lattice
plays the bulk portion in SRL, we can expect the bulk origin solution to be
identical to the solutions of Husimi lattice if the thickness is large enough
to ignore the effects of surface to bulk, and this expectation is confirmed in
the region $T<2$. However the 2-cycle bulk solutions differ from the Husimi
lattice solutions, and converge to $0.5$ solution quickly above $T=2$. Since
the bulk solution comes to be almost steady with thickness larger than $19$,
this difference is not really caused by the surface effect. Our calculation
requires a set of initial seeds for the recursive calculation as the procedure
described in Fig. \ref{fig15}, the bulk calculation takes the surface solution
as its initial seed, in this way, once the surface solution reaches $0.5$, the
initial seeds of $0.5$ will immediately affect the recursive calculation
inside the bulk and converges the bulk origin solutions to be $0.5$. The bulk
origin solutions at converging point with different thickness are shown in
Fig. \ref{fig24}. We can see the convergence occurs with very slight
differences no matter how large the thickness is, while theoretically we
should have the bulk origin solutions to be identical with Husimi bulk
solutions. Simple to say, because of the property of recursive calculation,
the bulk origin solution will be lead away from the exact description by the
effect of surface solution in the temperature region between surface melting
and Husimi bulk melting transition temperatures, which is not really the
\textquotedblleft surface effect\textquotedblright\ in nature.%

This is also a reason we are not interested in calculating the thermodynamics
of the whole SRL. Theoretically, the solution on each site of the SRL can be
approached and therefore we can calculate the thermal properties.
Nevertheless, there are three facts make this calculation unreliable. Firstly,
as mentioned above, in the temperature region between surface melting and
Husimi bulk melting transition, the bulk origin solution is affected by the
seeds of $0.5$ solutions on the surface. Secondly, the recursive calculation
technique requires several steps to reach the fix-point solutions. This
implies the calculations on the first few layers closing to the surface are
numerical instead of exact calculation. Although for large thickness bulk this
error can be neglected, the thermal behaviors associated with exact different
layers is not useful. Thirdly, the recursive structure of SRL proportionally
generates more surfaces with increasing thickness. Unlike the regular lattice,
in which the contribution of surface will be neglected with a sufficient large
bulk, the SRL will have half sites on the surface with infinitely large bulk
trees. In this way, the thermodynamics on each site is just the averaged value
of surface and bulk values. For a short summary, our approach of SRL is good
to track the thermal behaviors on the surface with the account of bulk
contributions, but not to discover the thermal behaviors change with various thicknesses.

\subsection{The transition temperatures reduction}

In section I we have reviewed the findings that the presence of a free surface
dramatically decreases the transition temperatures of bulk system. By
comparing the thermodynamics on the surface/thin film we achieved in section
II and III and in the bulk system, our calculations also clearly indicate the
reduction of both melting and ideal glass transitions temperature on the
surface. Fig. \ref{fig25} show the free energy comparison of Husimi bulk
system and SRL:%

\begin{figure}
[ptb]
\begin{center}
\includegraphics[width=0.8\textwidth]
{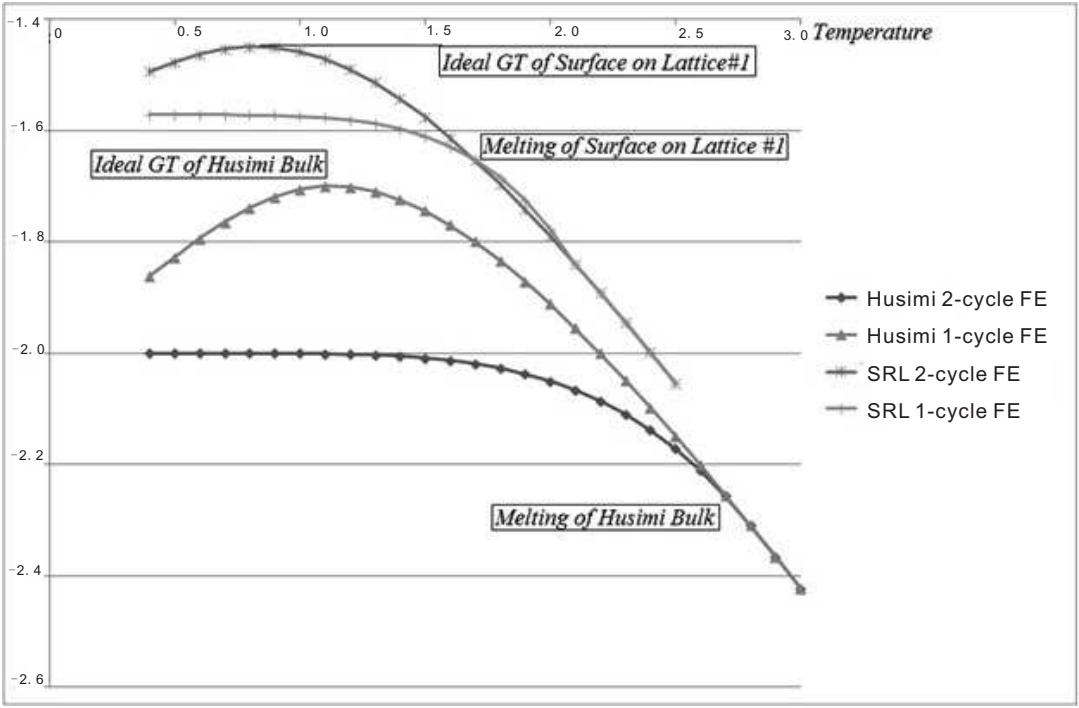}%
\caption{The free energy comparison of Husimi bulk system and SRL.}%
\label{fig25}%
\end{center}
\end{figure}

In SRL the melting and ideal glass transition temperature are dramatically
decreased comparing to Husimi bulk. In Section I we introduced the empirical
equation to describe the temperature reduction with the change of the
thickness. Since the thickness dependence of transition temperature reduction
is not available in our methods, and our calculations focus on the transitions
right on the surface, we may compare our results to the ratio of glass
transition temperatures of bulk and the thinnest free-standing film in others'
works, either experimental or simulation results.

The $T_{\text{K}}$ reduction ratio of SRL is $0.89/1.1=0.809$. In Forrest and
co-workers' work, the $T_{\text{g}}$ of the thinnest PS film they made is
$300$K and the $T_{\text{g}}$ of bulk PS is $369$K, the reduction ratio is
$0.813$ [10]. In Torres and co-workers' MD simulation, this reduction ratio of
a free standing film is $0.24/0.3=0.8$ [12]. In de Pablo and co-workers' MC
simulation, this ratio is $0.85/1.08=0.79$ [23].

Because our results are from the reference case, which is very simple without
any particular artificial parameters setup, the fact that our ratios are close
to others' results may only verify the validity or practice of our method. To
particularly describe a real system, the setup of energy parameters is the
most critical issue. The effects of energy parameters in our model will be
discussed in later section. The fact that similar reduction can be observed in
our small moleculars model implies that the lower transition temperature on
surface/thin film basically originates from the dimension reduction and less
interaction constraints. The chain-structure of polymer system may not be the
main reason for transition temperature reduction, although it does play
important roles to affect the transition properties.

\subsection{The effects of energy parameters}

In this section we are going to study the effects of different energy
parameters in SRL by the variation of one parameter with other parameters
fixed. The thickness is always set to be $19$ for providing a sufficient tree
size and relatively shorter calculation time consumption.

In Eq.\ref{surface_e_alpha} we specified the diagonal interaction energy
parameter on the surface to be $\overline{J_{\text{p}}}$, differed from the
diagonal interaction $J_{\text{p}}$ inside the bulk. And similarly we have
$\overline{J}$ , ($\overline{J\prime}$) and $\overline{J"}$. This
specification is to enable us setup different interaction circumstance on the
surface and provide more versatility in our model. However the effects of
either $J_{\text{p}}$ or its counterpart on the surface $\overline
{J_{\text{p}}}$ are the same. Thus in this section we are going to simplify
the case and setup $J_{\text{p}}$ and $\overline{J_{\text{p}}}$ , $J"$and
$\overline{J"}$ to be the same. Nevertheless, a variation with fixed
$\overline{J}$ will be discussed in details, because the nearest-neighbor
interaction $J$ and $\overline{J}$ have a much larger weight in the
Hamiltonian and play more critical roles in the system. The difference setup
of $J$ and $\overline{J}$ provides the simulation of surface tension which is
critical to the surface properties.

\subsubsection{The diagonal interaction $J_{\text{p}}$}

\begin{figure}
[ptb]
\begin{center}
\includegraphics[width=0.8\textwidth]
{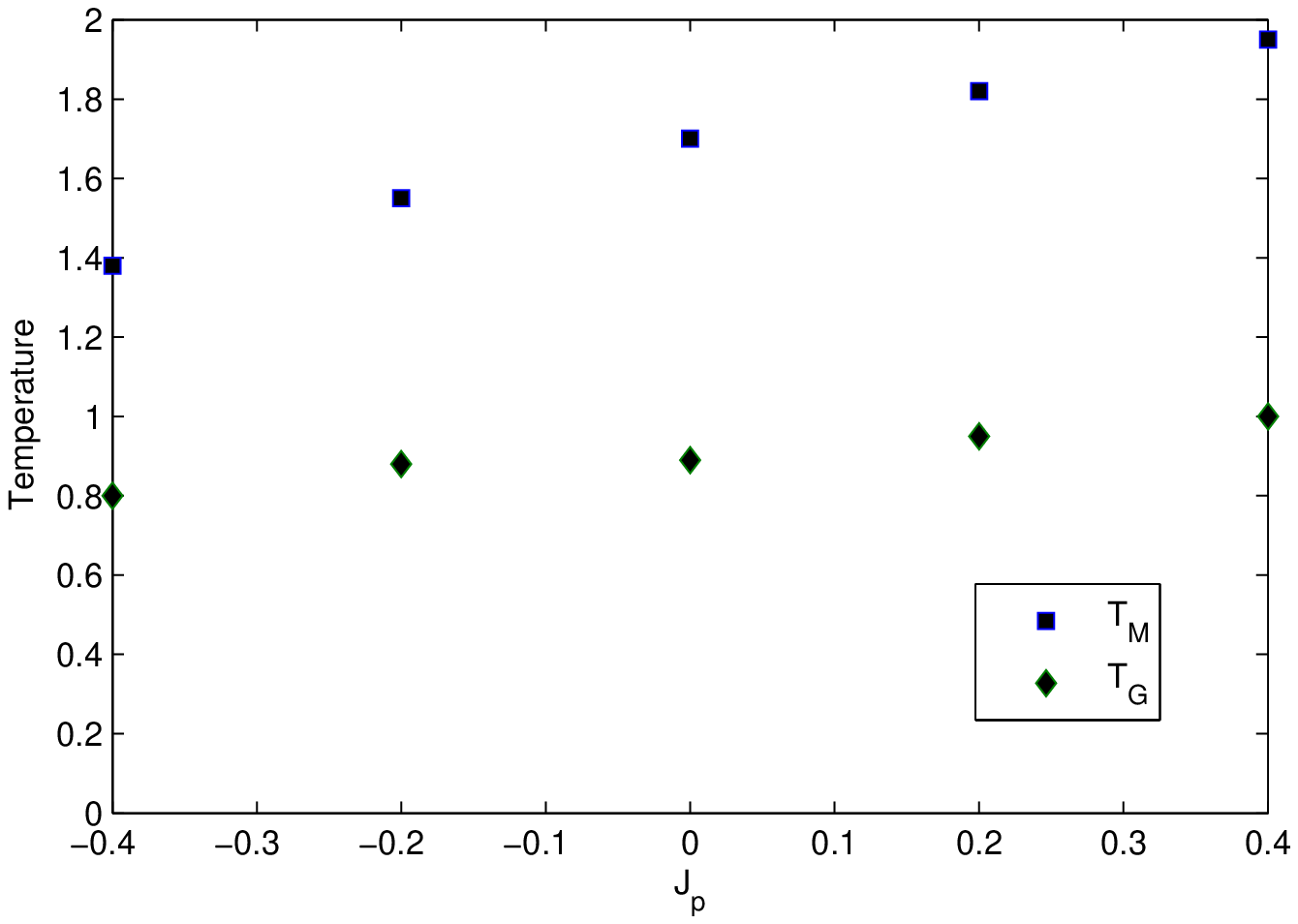}%
\caption{The transition temperature variations with different $J_{\text{p}}.$}%
\label{fig26}%
\end{center}
\end{figure}

In the Hamiltonian Eq.\ref{e_alpha}, the nearest-neighbor interaction $J$ is
negative by the definition of anti-ferromagnetic Ising model, the system
prefers an anti-aligned spins arrangement to obtain a lower energy with a
negative first term, then we can also observe that for an anti-aligned spins
arrangement, the diagonal spins pairs are in the same state. Therefore, unlike
the nearest-neighbor interaction $J$, a positive diagonal interaction will
make the second term to be negative, which encourages the alternating
arrangement and increases the transition temperatures (i.e. the crystal is
more stable to melt). On the other hand, a negative $J_{\text{p}}$ will
compete with $J$ and trend to destroy the ordered state, which decreases the
transition temperature. Because we have four nearest-neighbor and two diagonal
interactions in a square unit, the nearest-neighbor interaction outweigh the
diagonal interaction. Also, there is no diagonal interaction in the surface
single bond unit, which also lowers the contribution of $J_{\text{p}}$. The
variation of $J_{\text{p}}$ will only moderately change the transition
temperatures on the surface; the overall thermal behaviors are similar to the
reference case. The thermal behaviors with four different $J_{\text{p}}$
values $\pm0.2$ and $\pm0.4$ are calculated and the melting and ideal glass
transition change with $J_{\text{p}}$ variation is summarized in Fig.
\ref{fig26} and table 2. It is obvious that positive $J_{\text{p}}$ will make
the system more stable and increase the transition temperatures, or vice
versa. We also found that $J_{\text{p}}$ cannot be larger than $\pm0.5$
otherwise stable solutions can not be reached.%

%

\begin{tabular}
[c]{llll}%
\multicolumn{4}{l}{Table 2. The transition temperature variations with
different $J_{\text{p}}$}\\
$J_{\text{p}}$ & $T_{\text{m}}$ & $T_{\text{K}}$ & $T_{\text{m}}/T_{\text{K}}%
$\\
0.4 & 1.95 & 1 & 1.950\\
0.2 & 1.82 & 0.95 & 1.916\\
0 & 1.70 & 0.89 & 1.910\\
-0.2 & 1.55 & 0.88 & 1.761\\
-0.4 & 1.38 & 0.80 & 1.725
\end{tabular}

\subsubsection{ The quadruplet interaction $J"$}

\begin{figure}
[ptb]
\begin{center}
\includegraphics[width=0.8\textwidth]
{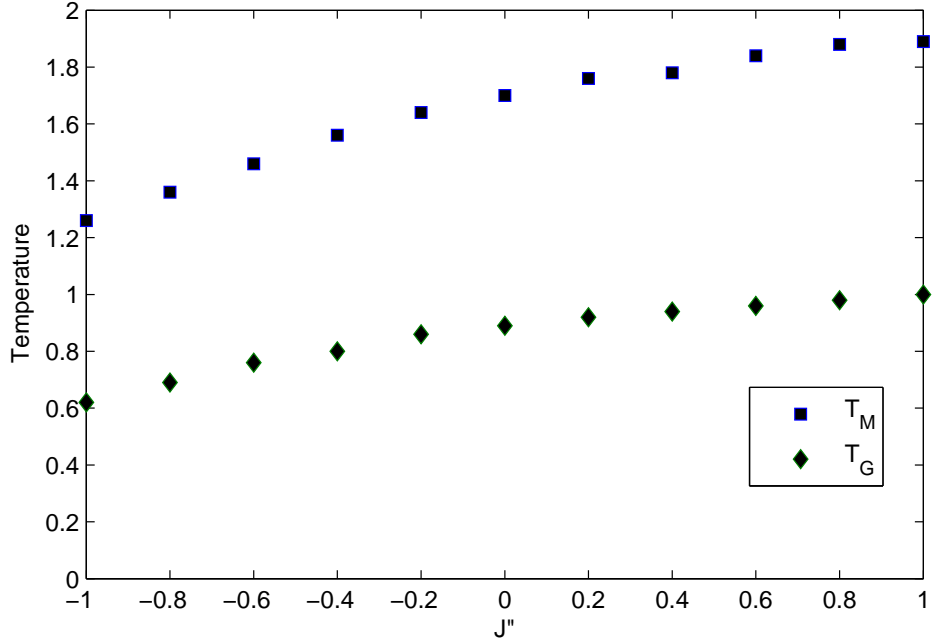}%
\caption{The transition temperature variations with different
J\textquotedblright.}%
\label{fig27}%
\end{center}
\end{figure}

The quadruplet interaction $J"$\ is a complicated term. By simply analyzing
the fourth term in Hamiltonian Eq. (2.1), it is difficult to give out a clear
expectation on the effects of $J"$\ since both system preferred or defective
structures can give out either positive or negative values of fours spins
product, then consequently either positive or negative values of $J"$\ can
against the crystallization. Transition temperatures of $J"=\pm0.2,\pm
0.4,\pm0.6,\pm0.8,$and $\pm1.0$ with other parameters fixed to be 0 are shown
in Fig. \ref{fig27}.%

From the graph above, we can see that the parameter $J"$\ has a similar effect
to Jp: positive J\textquotedblright\ will make the system more stable and
increase both transition temperatures, and vice versa. We also found that,
unlike the limited value of $J_{\text{p}}$, $J"$\ can be assigned with large
value such as $\pm1.0$ without destroying the stable solutions. The transition
temperatures variation with different $J"$\ are summarized in table 3.%

\begin{tabular}
[c]{llll}%
\multicolumn{4}{l}{Table 3. The transition temperature variations with
different $J"$}\\
$J"$ & $T_{\text{m}}$ & $T_{\text{K}}$ & $T_{\text{m}}/T_{\text{K}}$\\
1.0 & 1.89 & 1.00 & 1.89\\
0.8 & 1.88 & 0.98 & 1.92\\
0.6 & 1.84 & 0.96 & 1.92\\
0.4 & 1.78 & 0.94 & 1.89\\
0.2 & 1.76 & 0.92 & 1.91\\
0 & 1.70 & 0.89 & 1.91\\
-0.2 & 1.64 & 0.86 & 1.91\\
-0.4 & 1.56 & 0.80 & 1.95\\
-0.6 & 1.46 & 0.76 & 1.92\\
-0.8 & 1.36 & 0.69 & 1.97\\
-1.0 & 1.26 & 0.62 & 2.03
\end{tabular}

One important difference between the effects of $J_{\text{p}}$ and $J"$\ can
be observed in table 2 and 3: Although both parameters act similarly in
changing the transition temperature, with the $J"$\ variation the ratio of
$T_{\text{m}}/T_{\text{K}}$ is relatively constant unless $J"$\ is assigned
with extraordinary values like $\pm1.0$, while with the $J_{\text{p}}$
variation the ratio of $T_{\text{m}}/T_{\text{K}}$ also changes dramatically.
This difference could be useful in modifying the parameters setup to describe
the real systems or experimental results. For a real system, factors affecting
the thermodynamics can act in different ways, some of them may lift/reduce
both transition temperatures but keep the supercooled liquid region constant,
or some may change the window size of supercooled liquid state.

\subsubsection{The surface nearest-neighbor interaction $\overline{J}$}

\begin{figure}
[ptb]
\begin{center}
\includegraphics[width=0.8\textwidth]{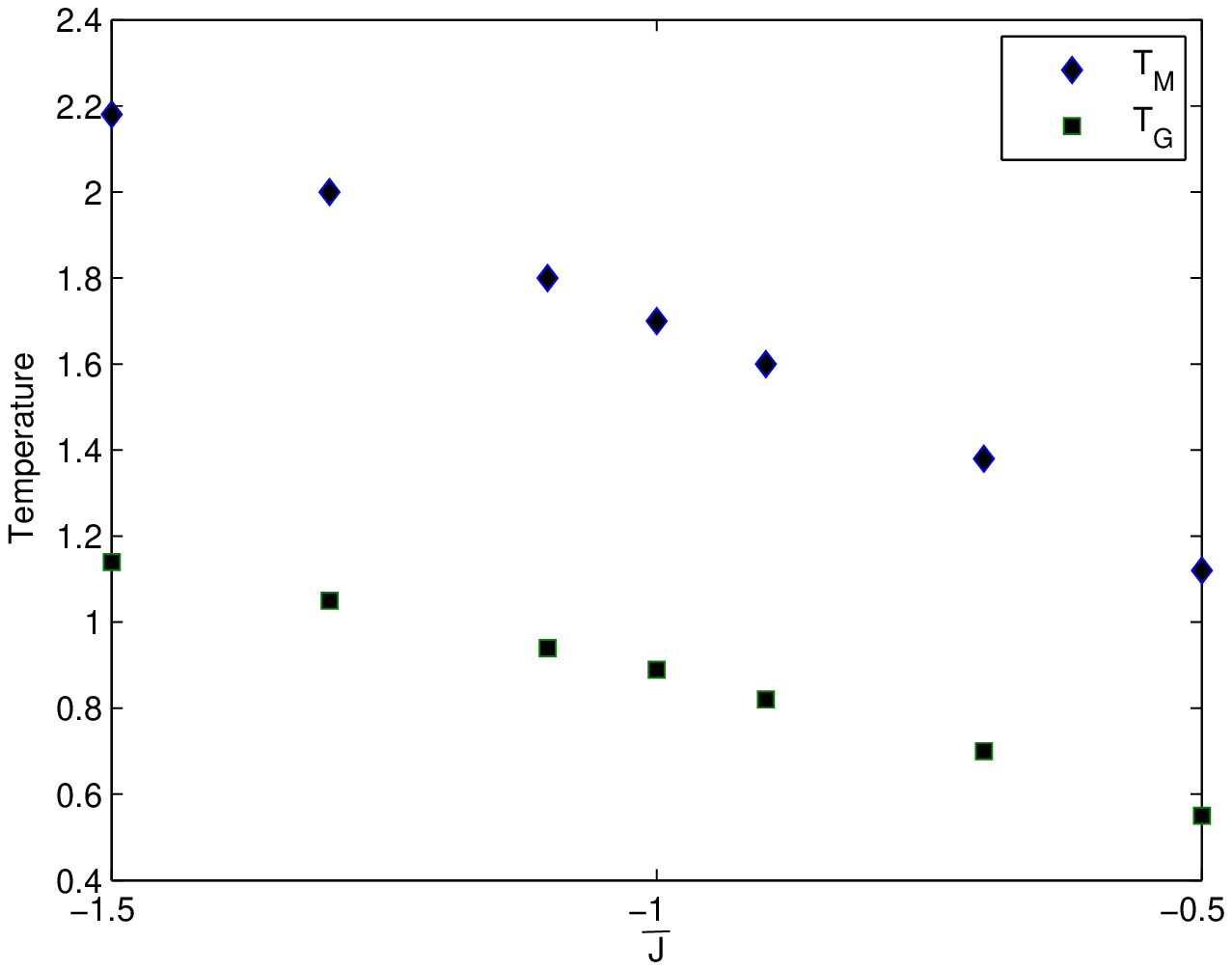}%
\caption{The transition temperature variations with different $\overline{J}$}%
\label{fig28}%
\end{center}
\end{figure}

As mentioned above, for convenience we setup the energy parameters in the bulk
and on the surface, such as $J_{\text{p}}$ and $\overline{J_{\text{p}}}$, to
be the same. Nevertheless, our lattice has a fixed bond length (this length is
unit $1$ in Ising model) between neighbor sites on the surface and in the
bulk, and we also know that the asymmetric interactions and the surface
tension are the main reason of unique properties of the surface. Therefore
with a fixed bong length, we have to somehow modify the surface interaction
circumstance to mimic the surface tension or asymmetric interactions. This can
only be made possible with different $J$ and $\overline{J}$ setup. Thus we
fixed the bulk nearest-neighbor interaction parameter $J$ to be $-1$, and
observe how the variation of surface bond interaction $\overline{J}$ affect
the thermodynamics on the surface. The transition temperatures change with
variation of $\overline{J}=-0.5,-0.7,-0.9,-1.1-1.3$ is summarized in Fig.
\ref{fig28} and table 4.

From Fig. \ref{fig28} we can conclude that larger absolute value of negative
$\overline{J}$ makes the system more stable and increases both melting and
ideal glass transition temperatures. The variation of $\overline{J}$ revealed
that the surface is more stable with large surface tension ($-\overline{J}%
>-J$), or easier to undergo transitions than in the bulk with small surface
tension ($-\overline{J}<-J$). Positive $\overline{J}$ was also been tested
however no stable solution can be reached. This is because the positive
$\overline{J}$ will prefer ferromagnetic-aligned spins arrangement which only
has $1$-cycle solution available. On the other hand it is also hard to imagine
a homogeneous system with particles attractive to each other in the bulk but
repulsive on the surface.%

%

\begin{tabular}
[c]{llll}%
\multicolumn{4}{l}{Table 4. The transition temperature variations with
different $\overline{J}$}\\
$\overline{J}$ & $T_{\text{m}}$ & $T_{\text{K}}$ & $T_{\text{m}}/T_{\text{K}}%
$\\
-0.5 & 1.12 & 0.55 & 2.04\\
-0.7 & 1.38 & 0.70 & 1.97\\
-0.9 & 1.60 & 0.82 & 1.95\\
-1 & 1.70 & 0.89 & 1.91\\
-1.1 & 1.80 & 0.94 & 1.91\\
-1.3 & 2.00 & 1.05 & 1.90\\
-1.5 & 2.18 & 1.14 & 1.91
\end{tabular}

\subsubsection{The triplet interaction J' and magnetic field H}

It is easy to understand that we have $0.5$ solutions at high temperature, and
2-cycle solutions symmetric to $0.5$ at low temperature, is because that the
magnetic field $H$ is $0$, the system has no bias on either $+$ or $-$ spins
on a particular site. Thus at high temperature one site has $50\%$ probability
to be occupied by either $+$ or $-$ spin. If a non-zero value of magnetic
field $H$ exists, we can expect the 1-cycle solution off the central $0.5$
line or asymmetric 2-cycle solutions. A positive $H$ will prefer more + spins
and moves the 1-cycle solution higher than $0.5$, or vice versa. In our
previous research on bulk Husimi lattice, the three spins interaction
(triplet) $J\prime$ acts similarly to magnetic field $H$ in changing the bias
of spins. However, in SRL the $J\prime$ and magnetic field $H$ are not useful,
although for a general presentation we still have them in our lattice model
and energy equations. The reason is that, the uneven property of the surface
on SRL determines the 1-cycle solution can only be $0.5$. If a non-zero
magnetic field $H$ or triplet interaction $J\prime$ presents, the 1-cycle
solution on the surface cannot be achieved. The single bond unit does not have
triple interaction $J\prime$ and the effect of $H$ is much smaller than it is
in the square unit, therefore the anti-aligning property of the single will
always prefer to have different spins on it. In another word, if a 1-cycle
solution can be achieved on the single surface bond, it must be the $0.5$
solution. Even we can still calculate the thermodynamics of 2-cycle solution
on the surface, we must have the corresponding 1-cycle solution to determine
the melting and ideal glass transitions. Therefore, we have to abandon the
variation of $H$ and $J\prime$, which is a disadvantage of the SRL.

\begin{center}
.
\end{center}

\section{CONCLUSION}

We applied recursive lattice technique to investigate the thermodynamics and
ideal glass transition on the surface/thin film of Ising spin system with a
theoretical base. A recursive lattice (SRL) were constructed to describe a
regular finite size square lattice, with a 2D bulk surrounded by 1D surface.
The SRL utilizes the finite-sized Husimi lattice, which is integrated by 2D
square units, to be the bulk part, and adopts single bond to connect the
finite-sized bulk parts. These single bonds and the outsider bonds of bulk
trees assemble the surfaces surrounding the bulk part and slipping through
independent bulk parts. The lattice is constructed with particular
approximations and compromises and is believed to be reliable approximation to
the regular square lattice in the aspect of coordination numbers, i.e. the
number of neighbor sites is $4$ inside the bulk and $3$ on the surface.

The Ising spins of two possible states $+$ or $-$ are put on the sites of SRL.
We assigned anti-ferromagnetic interaction in the model and different spin
states in neighbor pairs are preferred to be the stable state (crystal). The
partition function of a sub-tree with its base spin state fixed is defined as
partial partition function (PPF). Based on the recursive properties, the PPF
of one sub-tree can be expressed as a function of the PPFs of its
sub-sub-trees and a local weight. This recursive relation enables us to derive
the ratio of PFFs on one site. This ratio is called solution from which the
thermodynamics of system can be achieved by Gujrati trick. Two kinds of stable
solutions usually can be achieved by recursive calculation. One is in 2-cycle
form and represents the ordered state, the other is in 1-cycle form
representing the amorphous/metastable state.

The thermal behaviors of 2-cycle and 1-cycle solutions indicate the melting
transition by the cross point of free energies of two solutions, and the ideal
glass transition by the negative entropy of 1-cycle solution. Our results
agree with others' work that the transition temperatures are dramatically
reduced on the surface/thin film comparing which in the bulk. Nevertheless our
work shows that, regardless of specific properties of particular materials,
either polymer or small molecules, this transition temperature reduction can
be simply due to the dimension downgrade and lower interaction energy on the surface.

The effects of different interaction energy parameters other than
nearest-neighbor interaction such as diagonal interaction and four spins
(quadruplet) interaction were investigated. The variation of energy parameter
could either increase or decrease the stability of system and change the
transition temperatures according to the Hamiltonian. The behaviors of energy
parameters can be used to setup a particular model to describe the real system.

While our finding agrees well with others' experimental and simulation results, comparing to others' work, there are two advantages in our approach: 1) in this work our model focuses on
the small molecules system, it reveals the basic dimension origin of
transition temperatures reduction without involving the long chain properties
of polymer system; 2) the thermodynamics of systems are derived by exact
calculation method, the computation time is much shorter than typical
simulation methods, usually the calculation of one set of parameters in the
interesting temperature region can be done in less than 100 seconds.

\end{document}